\documentclass[12pt]{iopart}

\newcommand{\amplfi}{{\tt AMPLFI}}
\newcommand{\lalsim}{{\tt lalsimulation}}
\newcommand{\aframe}{{\tt Aframe}}
\newcommand{\mlgw}{{\tt ml4gw}}
\newcommand{\phenomd}{{\tt IMRPhenomD}}
\newcommand{\phenomp}{\tt IMRPhenomPv2}
\newcommand{\phenomxphm}{\tt IMRPhenomXPHM}
\newcommand{\raytune}{{\tt ray.tune}}
\newcommand{\pyro}{{\tt pyro}}
\newcommand{\nside}{{\mathrm{NSIDE}}}
\newcommand{\mc}{\ensuremath{\mathcal{M}}}
\newcommand{\msun}{\ensuremath{M_{\odot}}}
\newcommand{\mpc}{\ensuremath{\mathrm{Mpc}}}
\newcommand{\dl}{\ensuremath{D_L}}
\newcommand{\thetajn}{\ensuremath{\theta_{\mathrm{JN}}}}
\newcommand{\boldtheta}{\ensuremath{\mathbf{\Theta}}}
\newcommand{\data}{\ensuremath{\mathbf{d}}}
\newcommand{\boldgamma}{\ensuremath{\mathbf{\Gamma}}}
\newcommand{\smallgamma}{\ensuremath{\mathbf{\gamma}}}
\newcommand{\vicreg}{\ensuremath{\mathcal{L}_{\mathrm{VICReg}}}}

\usepackage{iopams}
\usepackage{graphicx}
\usepackage{url}
\usepackage[font=small]{caption}
\begin{document}

\title[AMPLFI]{Rapid Likelihood Free Inference of Compact Binary Coalescences using Accelerated Hardware}

\author{D. Chatterjee,$^{1,2}\;^{\dagger}$ E.~Marx,$^{1,2}$ W.~Benoit,$^3$ R.~Kumar,$^4$ M.~Desai,$^{1,2}$
E.~Govorkova,$^{1,2}$ A.~Gunny,$^{1,2}$ E.~Moreno,$^{1,2}$ R.~Omer,$^3$ R.~Raikman,$^{1,2}$ M.~Saleem,$^3$ S.~Aggarwal,$^3$ M.~W.~Coughlin,$^3$ 
P.~Harris,$^1$ E.~Katsavounidis$^{1,2}$}

\address{
$^1$Department of Physics, MIT, Cambridge, MA 02139, USA \\
$^2$LIGO Laboratory, 185 Albany St, MIT, Cambridge, MA 02139, USA \\
$^3$School of Physics and Astronomy, U. Minnesota, Minneapolis, MN 55455, USA \\
$^4$Department of Aerospace Engineering, IIT Bombay, Powai, Mumbai, 400076, India
}
\ead{$^{\dagger}$deep1018@mit.edu}
\vspace{10pt}

\begin{abstract}
We report a gravitational-wave parameter estimation algorithm, \amplfi, based on likelihood-free inference
using normalizing flows. The focus of {\amplfi} is to perform real-time parameter estimation
for candidates detected by machine-learning based compact binary coalescence search, \aframe. We present
details of our algorithm and optimizations done related to data-loading and pre-processing on accelerated
hardware. We train our model using binary black-hole (BBH) simulations on real LIGO-Virgo detector noise. Our model
has $\sim 6$ million trainable parameters with training times $\lesssim 24$ hours. Based on
online deployment on a mock data stream of LIGO-Virgo data, {\aframe~+ \amplfi} is able to pick
up BBH candidates and infer parameters for real-time alerts from data acquisition with a net latency of $\sim 6$s. 
\end{abstract}

\vspace{2pc}

\submitto{\MLST}

\section{Introduction}
It has been almost a decade since the discovery of gravitational waves (GWs) from compact binary mergers~\cite{AbEA2016a}, with
the last few years seeing a steady increase in the number of discovered GW events.
While the first observing run of the Laser Interferometer Gravitational-wave Observatory (LIGO) reported only three
events~\cite{AbEA2016a,AbEA2016g},~\footnote{GW151012, initially labeled as low-significance, was
later confirmed as a third event in O1.}
the number count stood at 90 within a span of five years~\cite{LSC2021djp}. Furthermore, the current ongoing fourth
observing run (O4) of ground-based observatories LIGO/Virgo/KAGRA (LVK) has already reported more than one hundred events discovered
online.~\footnote{See \url{https://gracedb.ligo.org/superevents/public/O4/} for the most updated list.} The trend is expected to continue with
the instrument getting closer to design sensitivity in fifth observing run.\footnote{See
\url{https://observing.docs.ligo.org/plan/} for observing plans.}

In parallel, the scope of multi-messenger astronomy (MMA) with GWs has seen a steady increase in terms of
effort and infrastructure being invested for the joint follow-up of GW signals with EM and other high-energy astrophysics counterparts. 
The online alert infrastructure of the LVK currently reports GW discoveries along with follow-up data products in $\sim 30$s
after merger time~\cite{Chaudhary_2024}. Early-warning searches~\cite{Cannon_2012} that can potentially pick up low-mass
BNS systems up to $\sim 1$ minute before merger have been deployed online~\cite{Magee_2021}. 
The alert distribution mechanisms, like NASA GCN,\footnote{\url{https://gcn.nasa.gov/}} have
seen upgrades~\cite{2022GCN.32419....1B}, and new alert brokers like SCiMMA\footnote{\url{https://scimma.org/}} have
become available for the community to use. Publicly available services like TreasureMap~\cite{2019GCN.26244....1W}
have seen a steady adoption from observatories to share observed and scheduled fields to orchestrate observations. 
Tools like SkyPortal~\cite{skyportal2019} and TOM-Toolkit~\cite{tom_toolkit} have been developed
to aid target-of-opportunity followup. 

All this development comes at a time when the number of GW discoveries have significantly increased
corresponding to the improvement in sensitivity of Advanced
LIGO~\cite{PhysRevD.102.062003,aLIGO}, Advanced Virgo~\cite{VIRGO:2014yos}, and KAGRA~\cite{10.1093/ptep/ptaa125}
instruments, and the sensitivity of time-domain telescope facilities allow for unprecedented discovery
rates. However, identifying gravitational-wave counterparts jointly have been extremely challenging. The discovery of
GWs and multi-wavelength  EM emission from the merger of the binary neutron star (BNS), GW170817,~\cite{gw170817,gw170817_mma}
remains the first and only success story, albeit a rarity, with most subsequent candidates likely to be at much farther
distances~\cite{2020NatAs...4..550C}. 

One primary step toward improving follow-up campaigns is the availability of fast, real-time Bayesian parameter
estimation (PE) of compact binary coalescence (CBCs) to provide accurate data products for GW follow-up. The
computationally expensive part of stochastic sampling techniques, like nested sampling currently in use, involve the
repeated computation of the likelihood. Techniques like reduced-order-quadrature (ROQ)~\cite{Canizares:2014fya,morisaki2023rapidlocalizationinferencecompact}
and focused-ROQ~\cite{Morisaki:2020oqk} have been developed in view of making real-time PE as fast as possible. This
is currently used in LVK to deliver update alerts from Bayesian parameter estimation on the timescale of several
minutes to hours. Other techniques like the use of accelerated hardware for stochastic sampling has been reported in
\cite{Talbot:2019okv, wong2023fastgravitationalwaveparameter}, and mesh-free approximation for
sky-localization, reported in \cite{PhysRevD.108.064055,PhysRevD.109.024053}.

More recently, likelihood-free inference (LFI), using variational methods, have emerged as a different paradigm
with flexible neural network approximators being used to learn the posterior or the likelihood. Their use has
been demonstrated on GW data~\cite{Gabbard_2021,PhysRevD.108.042004}, in particular with posterior
estimation using normalizing flows~\cite{Dax_2021}, such as in the DINGO algorithm. 
However, in order to relay discovery alerts for prompt followup, the combination of \emph{search
and inference} needs to be considered together.\footnote{This is done in case of stochastic signals offline,
for example, see \cite{Smith_2018}.}
Also considering a live, real-time system design, several overhead
costs like data transfer, file input/output operations, communicating data to a remote server, and so on are often
overlooked in isolated analyses, but show up in the overall time-to-alert.
It is also worth highlighting that traditionally in GW data analysis, the search and PE components have been treated
separately -- match-filtering searches pick up the candidates from the data
stream using suitable detection statistic, but also provide important context like the best matching template and the signal-to-noise
ratio time series, which is then used to compute sky-localization maps~\cite{Singer_2016} and EM-bright source
properties~\cite{Chatterjee_2020} sent out in the sub-minute alerts by the LVK~\cite{Chaudhary_2024}. The results are then
updated based on Bayesian PE results, in few hours timescale.

Although machine-learning techniques like LFI bring
promise, large model size and/or long training times can be a barrier for operations.
Also, given the slowly changing background over the course of days to weeks, the algorithm
should be re-trainable from a previous model state, preferably without investing on expensive and dedicated online
hardware for this purpose. This is currently lacking for online models like DINGO, which report 10-day training time
on a NVIDIA A100 GPU~\cite{Dax_2021}.

In this work, we try to address the points highlighted above in the context of fast online search and parameter estimation
for MMA with GW. We report \amplfi,\footnote{\textbf{A}ccelerated \textbf{M}ultimessenger \textbf{P}arameter estimation using \textbf{LFI}; pronounced ``amp-li-fy''}
a PE algorithm based on LFI using normalizing flows.
The primary focus of \amplfi~is to run alongside neural-network based CBC search {\aframe}~\cite{aframe}, and compute
GW alert data products like skymaps and other use source-properties to be sent out with LVK discovery alerts. Though the core
principles of LFI and its application are similar to efforts mentioned above, the technical implementation is independent and
focused toward online inference. In particular, there are several elements of GW data analysis that are re-implemented
as a part of {\mlgw} (codebase: \url{https://github.com/ML4GW/ml4gw}), designed for running on accelerated hardware like GPUs
for fast and efficient training and inference. Some common set of tools from {\mlgw} are used by both {\aframe}
(codebase: \url{https://github.com/ML4GW/aframev2/}) and {\amplfi} (codebase: \url{https://github.com/ML4GW/amplfi}),
the latter being the focus of this work. 

We outline the
rest of the paper as follows. In section~\ref{sec:philosophy}, we motivate our design principles toward running search
and PE together. In section~\ref{sec:simulations}, we mention optimizations related to data pre-processing and implementing
simulations on accelerated hardware which ensure that most of the computation is occurring on the GPU. In
section~\ref{sec:embedding-net}, we present the details of a data embedding network which is used to
summarize the data. This embedding is pre-trained using a self-supervised method to create data summary marginalizing
parameters that are not of interest (at least from GW alerts standpoint). In section~\ref{sec:estimation}, we
give the details of our normalizing flow implementation. We present results and benchmarks in section~\ref{sec:results},
before concluding in section~\ref{sec:conclusion}.

\section{\aframe~+ \amplfi}\label{sec:philosophy}
In order to build as fast of a system as we can, we have made a number of design choices  when building the {\aframe~+ \amplfi}
framework that we highlight in the following:
\begin{itemize}
    \item A modular design to perform search and PE. The search for GW signals in this case is done by {\aframe}.
          Candidates from {\aframe} provide an estimate of the time of arrival and the significance via a false-alarm-rate (FAR).
          This is unlike traditional match-filtering searches that provide, in addition, the best matching template, and the
          corresponding signal-to-noise time series that is used by other annotation algorithms to provide 
          skymaps~\cite{Singer_2016} and source properties~\cite{Chatterjee_2020} of binary systems.
          In the proposed framework, once a segment of data is found to be of high-significance i.e., containing a GW signal,
          the parameters are inferred using {\amplfi}. Hence, Bayesian parameter estimation results are available
          along with the discovery of the candidate.
    \item Data is held in GPU memory to minimize overheads in communication between different components in the low-latency alert
          infrastructure~\cite{Chaudhary_2024}. For example, {\aframe} runs as a service, maintaining a
          buffer~\footnote{A snapshotter that only sends new data segments into GPU memory} of the data in GPU
          memory. Once a trigger occurs, the relevant segment is passed to {\amplfi} for inference. Based on model size
          of {\aframe} and {\amplfi}, both are able to be served on a single GPU like NVIDIA A30. This reduces any inference
          overheads as the data is kept on the same device. However, the models may be served as separate micro-services
          in case the model size or running on a single GPU turns out to be a barrier.
    \item The accuracy of the results are suited for ``online'' purposes i.e. the aim is to provide
          data products like skymaps, and source properties for online LVK alerts. Therefore, we restrict
          ourselves to GW waveform models that capture the inspiral-merger-ringdown phases, but do not focus on physics
          of higher-modes, spin precession etc., and prioritize fast inference for data products required for follow-up.
\end{itemize}

\noindent For \amplfi~ we use a normalizing flow to learn the posterior distribution directly using simulations of binary
black hole signals (BBHs). We also make some optimizations compared to previous efforts
in light of an online inference algorithm:
\begin{itemize}
    \item We use real detector data from the LIGO GW instruments during training. Most previous efforts use simulated,
          colored Gaussian noise.
    \item We use efficient data loading and whitening tools to minimize the data transfers back
          and forth between CPU and GPUs (or other accelerators). We elaborate this below in section~\ref{sec:simulations}.
    \item We re-implement CBC waveform generation on the GPU memory to directly generate waveforms on-the-fly
          during training.
\end{itemize}

\section{Simulations on Accelerated Hardware}
\label{sec:simulations}
\subsection{Waveform model}
\label{sec:phenomd}
\begin{figure}
    \centering
    \includegraphics[width=0.8\textwidth]{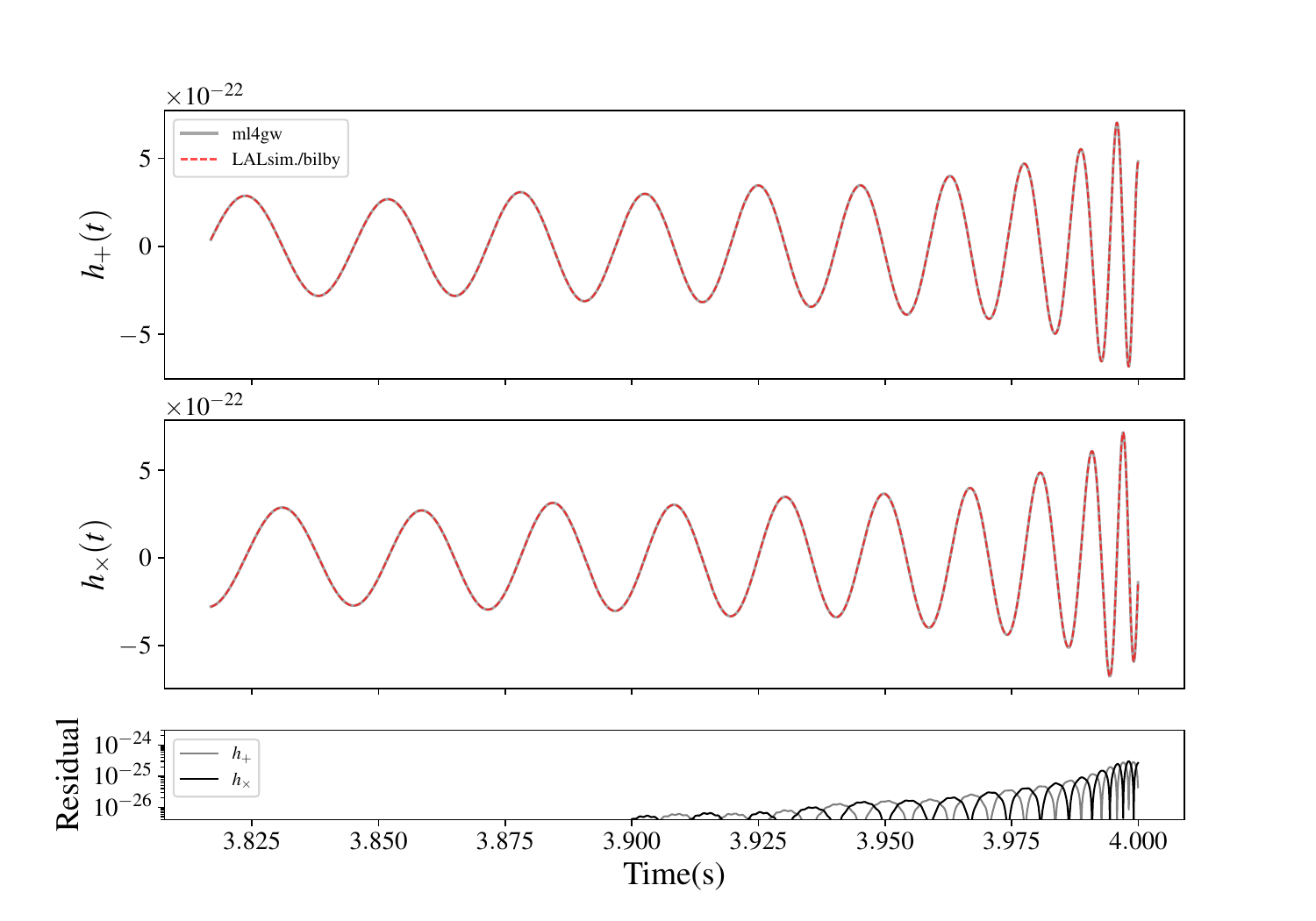}
    \caption{A comparison of time-domain {\phenomd} between that implemented in this study, as a part of the
    {\mlgw} library. The parameters of the waveform is ${\mc = 26\;\msun, q = 1.0, \dl = 1000\;\mpc, \chi_{1,2} = 0.0}$.
    We find that while there are differences in the waveform strain, the residuals are below three orders of magnitude
    compared to the signal for most of the evolution, except the final few cycles where it is two orders of magnitude
    lower.
    }
    \label{fig:waveform}
\end{figure}
We use the \phenomd~phenomenological waveform model for our simulations~\cite{Khan:2015jqa}. This waveform model contains
the full inspiral-merger-ringdown physics, starting with the inspiral phase up to 3.5 post-newtonian order in GW phase
(known as TaylorF2; see~\cite{Buonanno_2009} for a review), and using an ansatz for the merger and ringdown, fitting them to numerical relativity
results. One limitation of this waveform model is the restriction to aligned spins i.e. BH spin components perpendicular
to the orbital plane and therefore no precession. Current online PE using stochastic sampling techniques use
the {\phenomp} waveform approximant, which contains precessing effects. Furthermore, high mass BBH systems use the
{\phenomxphm} approximant, which also includes higher modes of radiation. However, we note that inference like sky-localization
is insensitive to such effects. Also EM-brightness of a binary depends primarily on the aligned spin
components aside from the mass ratio. Hence, the use of aligned-spin is justified for online purposes. In the future,
however, we plan to implement and integrate the {\phenomp} waveform model with our workflow.

In Figure~\ref{fig:waveform}, we show the time-domain strain of a representative BBH system based on our \phenomd~
implementation and compare it with that implemented in \lalsim. The latter is a part of the LIGO Algorithm Library~\cite{lalsuite},
and provides the core components of GW data analysis with LVK data. We find consistency between our implementation and that
in ~\lalsim, with the residual errors below the signal by a couple of orders of magnitude throughout the evolution.
Such residuals are unlikely to impact the results since the statistical errors of posterior is greater than systematic
error due to such differences. We present some comparison results in section~\ref{sec:results}.
Though we show comparison with a single representative system in Figure~\ref{fig:waveform}, several other combination
of parameters are tested for consistency with {\lalsim} as a part of unit-tests of the {\mlgw}
codebase (\url{https://github.com/ML4GW/ml4gw}).

\subsection{Data generation on the GPU}
\begin{figure}
    \centering
    \includegraphics[width=0.85\textwidth]{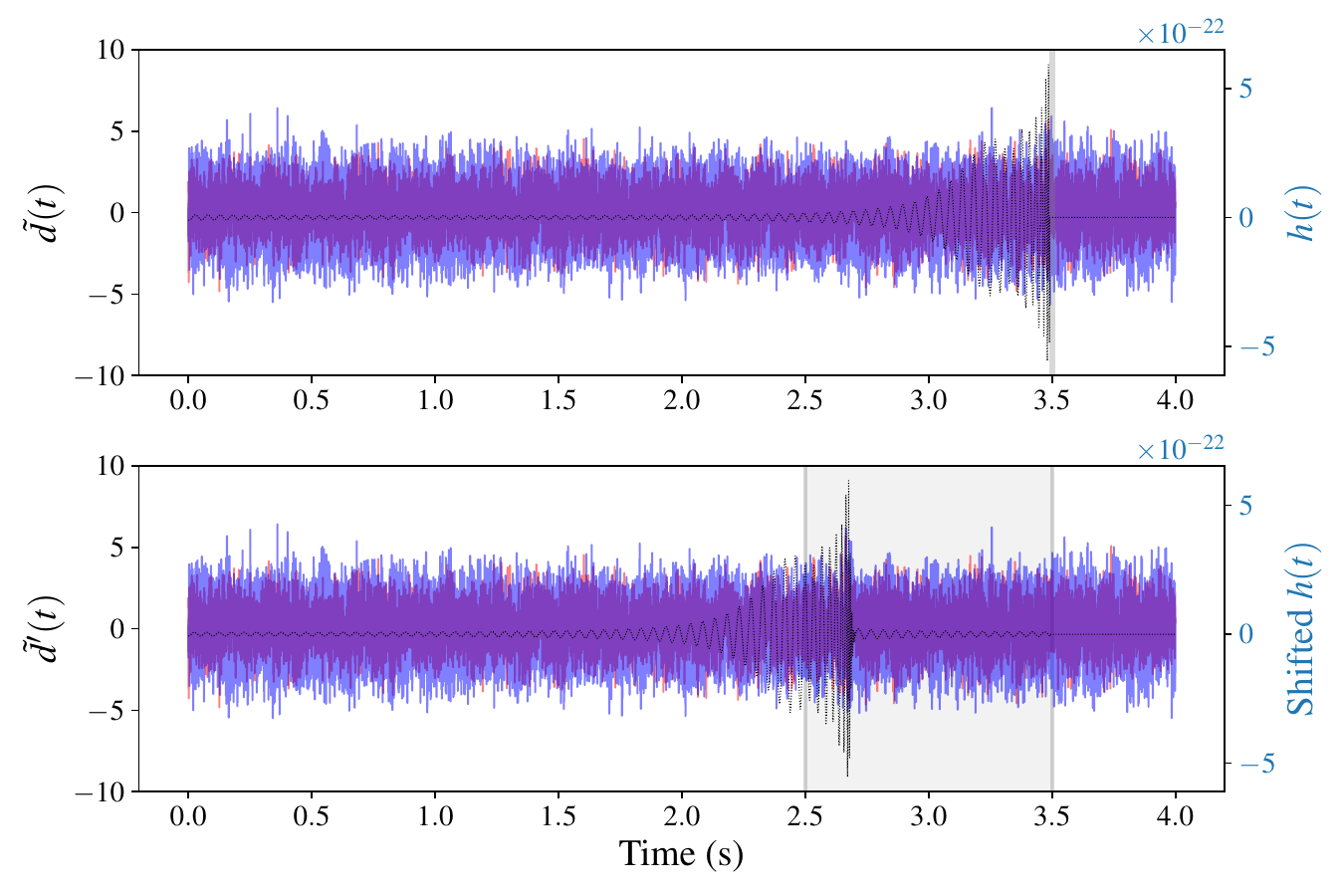}
    \caption{The figure shows the whitened time-domain background strain from Hanford and Livingston (HL) in two
    different colors. This is a stretch of data from May 2019 (early O3). A simulated BBH signal is injected;
    the waveform is overlayed. In the bottom panel, we see the same background strain with a time-shifted
    signal injected, the parameters of which are otherwise the same as the top panel. The time shift
    is random up to 1s compared to the top panel, indicated by the shading. We summarize our
    data views like $d$ and $d^\prime$, and embedding them jointly. Subsequently, for LFI, we
    only use $d^\prime$-like views.}
    \label{fig:data-view}
\end{figure}
Generally, neural-network models are trained using batches (also called mini-batches) of data that is pre-processed on the CPU
and then transferred to the GPU (or other co-processor) to carry out the forward/backward passes, and updating the model weights.
However, this may leave the GPU under utilized if the
pre-processing and data transfer between the CPU/GPU takes greater time compared to the operations to train the model. This is especially
important in the context of LFI since efficient training relies on providing unique combinations of
parameters and data to approximate the distribution. We therefore take a different approach by performing the data
generation on the GPU which allows generation of batches of data on the GPU, which is faster and at the
same time avoids the data transfer overheads. Additionally, all data pre-processing, like fourier transforms, power spectra estimation,
data whitening, are carried out on the GPU. This ensures consistent GPU utilization. Also, we can take advantage of the fact that
GPU architecture today provide large memory. Our workflow involves:
\begin{itemize}
    \item Transferring a chunk of two detector (Hanford and Livingston, subsequently HL) time-domain strain data, typically quarter of a
          day, sampled at 2048 Hz to the GPU before commencing training. The power spectra is fit to this background chunk.
          During training, $N$ small background chunks, each of 4s duration, are lazily loaded from the total training chunk,
          where $N$ is the training batch size.
    \item We generate $N$ points from our parameter prior and generate the \phenomd~waveforms directly on the GPU, as mentioned in
          section~\ref{sec:phenomd}. This step is fast, for example, generating $N=1000$ waveforms $\sim 0.15$s on a NVIDIA A40 GPU. 
    \item We then inject the signals into the background chunks, obtaining the data batch. We whiten the batch using the estimated
          power spectra and pass it along with the parameters for training/validation/testing. Examples of the whitened data with
          the injected waveform overlayed is shown in the panels of Figure~\ref{fig:data-view}.
\end{itemize}
We call this implementation {\tt InMemoryDataset} in \mlgw. We also note that though we have used GPUs as the co-processors in this work,
our software framework can also be ported to other accelerators, like TPUs or HPUs, supported by the {\tt{pytorch-lightning}}~\cite{falcon_2024_10779019}
framework that we use.

\section{Embedding Network}
\label{sec:embedding-net}
Our input to the neural network model is a 2-channel Hanford-Livingston (HL) 4s whitened time-domain strain. This
is projected into a lower-dimensional representation using a embedding network before performing LFI.
The coherent analysis of both channels of data is important for some aspects of GW parameter estimation,
like sky-localization since it depends on the difference in time of arrival in the different
instruments. Our embedding network resembles the ResNet architecture used in \aframe.
The implementation closely resembles that of the \texttt{torchvision} library, with some differences.
Firstly, {1-D} convolutions are used for time-series, instead of {2-D} variants used for images. We use group
normalization~\cite{wu2018groupnormalization} instead of batch-normalization. Also,
in \aframe, the architecture closely resembles a 34-layer residual network~\cite{he2015deepresiduallearningimage}.
We, however, avoid the final 512-channel stack of convolution blocks (see Figure.~3 in ~\cite{he2015deepresiduallearningimage})
since we do not find performance improvement after including the same. Thus our layer stacks contain blocks
of 64, 128, and 256-channel residual convolution blocks. The number of convolution layers in each block is determined
by hyper-parameter optimization (HPO) using Variance-Invariance-Covariance Regularization (VICReg)~\cite{bardes2022vicreg} loss,
detailed below in section~\ref{sec:ssl}.
Details about the HPO are presented in~\ref{appendix:embedding-hpo}. The best configuration resembles an
analogous 24-layer ResNet. A final fully-connected layer projects to an representation dimension
$D_{\gamma}=8$, which is also determined as a part of the HPO.

\subsection{\label{sec:ssl}Self-supervised learning of nuisance parameters}
We pre-train the embedding network to marginalize over uncertainties in arrival time up to 1 second.
This is done since the \emph{peak} of the detection statistic reported by {\aframe} may differ from the
true arrival time up to tens of milliseconds. We choose 1-second as a conservative upper bound for the same.
The pre-training is done via self-supervised learning (SSL) by identifying two ``views'' of the data as
being the similar, and training the embedding network to minimize VICReg. Examples of two different views,
$d$ and $d^\prime$, as shown in Figure~\ref{fig:data-view}, where the upper panel shows a signal that
is injected at a fixed reference time, while the lower panel shows a time-shifted signal i.e. all signal
parameters being the same except the time of arrival, which is chosen randomly up to 1s in this case.
Two batches of views are then forward-modeled through the embedding network and projected down to the resulting space. 
This projection, $\boldgamma$, is performed in two different steps via the ResNet, $f$ mentioned above,
which projects the inputs in to an 8 dimensional space $\smallgamma\in\mathbb{R}^8$, then expanding this
projected space using another fully-connected network, $h$, which takes $\gamma$ to a 24-dimensional space
$x\in\mathbb{R}^{24}$. The resulting composition is given by  $\boldgamma \equiv h\circ f$,
\begin{equation}
    \smallgamma = f(\data);\;\smallgamma' = f(\data');\;x = h(\smallgamma);\;x' = h(\smallgamma').
    \label{eq:representation}
\end{equation}
We follow the prescription mentioned in~\cite{bardes2022vicreg}
and compute the \vicreg~loss in the expanded dimension as,
\begin{eqnarray}
    \vicreg(x, x') &=& \lambda_1\;\mathrm{MSE}(x, x') +
      \lambda_2\;\left[\sqrt{\mathrm{Var}(x) + \epsilon} + \sqrt{\mathrm{Var}(x') + \epsilon}\right] + \nonumber \\
      &&\lambda_3\;\left[C(x) + C(x')\right].
  \label{eq:loss}
\end{eqnarray}
Here, $\mathrm{MSE}$ is the mean-squared error between the two projected views. The second term involves the
variances of the individual batches, regularized by a tolerance to prevent collapsing to zero. Finally, the
third term is the quadrature sum of the off-diagonal entries in the individual covariance matrices of the views.
The $\lambda_{1,2,3}$ are relative weights of each term, which we also select after hyper-parameter tuning.

Previous work using LFI reported other techniques in particular, group equivariance, to tackle this~\cite{dax2023group}.
We, however, take a different approach since parameters like time of coalescence are not as important for
follow-up as masses and sky location of the signal, and marginalize them in our data summary. For more details
on this technique, the reader is referred to~\cite{chatterjee2023optimizing}. This technique can be extended to
other nuisance parameters in case of GWs, for example the coalescence phase. This is left to future work.

\section{Posterior Estimation}
\label{sec:estimation}

\begin{table}
\caption{Prior distributions of parameters. Note that the distance prior is
a power-law with index 2 i.e. uniform in volume; cosmological effects are not
included.}
\centering
\begin{tabular}{ll}
\br
Parameter & Prior \\
\br
$\mc$ & Uniform(10, 100)$\msun$ \\
$q$ & Uniform(0.125, 1) \\
$\dl$ & Uniform in Vol.(100, 3000) $\mpc\;(\sim\dl^2)$ \\
$\thetajn$ & Sine(0, $\pi$) \\
$\alpha$ (RA) & Uniform(0, $2\pi$)\\
$\delta$ (Dec.) & Cosine($-\pi/2$, $\pi/2$) \\
$\phi_c$ (Coal. phase) & Uniform(0, $2\pi$) \\
$\psi$ (Pol. angle) & Uniform(0, $\pi$)\\ 
\hline
\end{tabular}
\label{tab:priors}
\end{table}

Posterior estimation in LFI involves learning the posterior, $p(\boldtheta \vert \data)$, using an approximator,
$q_{\varphi}(\boldtheta \vert \data)$, using simulations $\{\boldtheta_i, \data_i\}$. The parameters $\varphi$ are
adjusted to maximize the likelihood of the simulations, which is mathematically equivalent to minimizing the
Kullback-Leibler (KL) divergence between the true posterior and the approximator. The loss function used is,
\begin{equation}
    - \ln \mathcal{L}(\varphi) = - \frac{1}{N_{\mathrm{sims.}}}\sum_{i \in \mathrm{sims.}} \ln q_{\varphi}(\boldtheta_i \vert \data_i),
    \label{eq:log_likelihood}
\end{equation}
where the simulations are forward modeled to calculate their likelihood, which is then maximized during training.
The density evaluations are done by learning a set of variable transforms that take the original variables $\boldtheta$
to variables of a simpler base distribution, like a standard normal, which we use here. Several techniques are used
to build flexible transforms and preserve the probability density at each stage. We refer the reader
to a review article on normalizing flows and the different implementations~\cite{papamakarios2021normalizing}.
In our case the parameter space is 8-dimensional,
\begin{equation}
    \boldtheta = \{\mc, q, \dl, \thetajn, \alpha, \delta, \psi, \phi_c \}.
    \label{eq:parameters}
\end{equation}
The prior distribution used for generating the simulations is mentioned in Table~\ref{tab:priors}.
Note that the time of coalescence is not a part of parameter set since it is marginalized over.
Although the \phenomd~supports BH spins, we have ignored it for the current work. This will
be relaxed in subsequent versions of the code and reported in a future work. The data, \data,
consists of 4s sampled at 2048 Hz of whitened time-domain strain containing a BBH signal, as illustrated in
Figure~\ref{fig:data-view}. When training the normalizing flow, we condition the parameters on
the data representation, \smallgamma, as shown Eq.~(\ref{eq:representation}). This uses
only the ResNet, $f$. The expander, $h$, is not required subsequently. Also, we
leave the weights of $f$ to change further as a part of training the normalizing flow. Hence,
our normalizing flow maximizes,
\begin{equation}
    - \ln \mathcal{L}(\varphi) = - \frac{1}{N_{\mathrm{sims.}}}\sum_{i \in \mathrm{sims.}} \ln q_{\varphi}(\boldtheta_i \vert f(\data_i)),
    \label{eq:log_likelihood_mod}
\end{equation}
where the difference between Eq.~(\ref{eq:log_likelihood}) vs. Eq.~(\ref{eq:log_likelihood_mod}) is
the conditioning on the data summary, pre-trained to marginalize time of arrival. It should be
mentioned that the pre-training step with VICReg loss is significantly cheaper compared to training
the normalizing flow. In our experiments, we found the pre-training requiring few-tens of epochs
of training, which took less than an hour to reach early-stopping condition  on a NVIDIA A40 GPU.

\subsection{\label{sec:normalizing-flow}Autoregressive Flows for LFI}
Our normalizing flow implementation uses inverse auto-regressive transforms~\cite{kingma2017improving}.
This kind of auto-regressive transforms can be sampled in one forward pass. However, evaluating the
density requires $D$-forward passes, where $D$ is the dimensionality of the parameter space i.e. $D=8$
in our case. Masked auto-regressive transforms~\cite{papamakarios2018masked} on the other hand use similar
masked linear layers~\cite{pmlr-v37-germain15}, but on the contrary the density evaluation takes a single
forward pass and sampling is $D$-times as expensive. Although auto-regressive flows are \emph{universal
approximators}~\cite{papamakarios2021normalizing}, our choice of using inverse auto-regressive 
flow (IAF) is because our inference requires fast sampling, which is achieved in
a single pass with the IAF. In addition to IAF, coupling transforms were also considered. However,
in our experiments, we found such transforms to perform less optimally when constrained to the same number of trainable parameters.

Our transforms are implemented using the open source library {\pyro}~\cite{JMLR:v20:18-403}. The complete
transform is composed of 60 individual affine-autoregressive transforms. Each transform has 6 masked linear layers;
each layer having 100 hidden units. More complex transform functions like monotonic splines and neural-network with
positive weight exist in the literature. However, we use the affine transforms for simplicity and lower
number of trainable parameters. We train the network with a batch size of 800, with 200 batches per epoch
using the AdamW optimizer~\cite{loshchilov2019decoupledweightdecayregularization} with initial learning rate of
$1\mathrm{e}{-3}$ and weight decay of $2\mathrm{e}{-3}$. The learning rate is scheduled to reduce by a factor of 10
upon plateauing of the validation loss with a patience of 10 epochs. These configurations are decided after
hyper-parameter tuning over a combination of several hundred parameter combinations detailed in \ref{appendix:flow-hpo}.
Our trainer is scheduled to terminate training once the validation
loss saturates with a patience of 50 epochs. In terms of training time, we find training $\gtrsim 200$
epochs with the above configuration takes $\sim 20-24$ hours depending on a single 40GB NVIDIA A40/40GB NVIDIA A100
GPU. We note that because our dataset is generated on-the-fly, distributed training does not provide any
across multiple devices. We did not find significant differences in model performance by training across one vs.
multiple devices. In terms of number of trainable parameters, the embedding network contains $\sim 2.6$ million
parameters, while the auto-regressive transforms contain $\sim 3.2$ million parameters, totaling to $\sim 6$ million
parameters. We would like to note that with on-the-fly data generation on the GPU, a typical training run sees
$\gtrsim 200\;\mathrm{epochs} \times 200\;\mathrm{batches\;per\;epoch} \times 800\;\mathrm{batch size} \sim 32$
million unique simulations. This implies that unlike most ``large'' neural-networks in the literature today,
our network is not over-specified in the sense that training data samples exceed the number of trainable
parameters by several factors.

In terms of inference, average sampling time for drawing 20,000 posterior samples, conditioned on new data,
takes $\sim 0.05$s on NVIDIA A40 GPU. The same on a Intel Core i7
with 16 cores, takes $\sim 1-2$s. However,
it should be noted to create a sky-localization map in the HEALPix format~\cite{GoHi2005}, used in the GW
data analysis, the density needs to be evaluated across all pixel coordinates. We find the average
time to drawn 20K samples, and then evaluate the density across all pixels on the sky for a HEALPix resolution
of $\nside=32$ is $\sim 0.6$s. This is due to the choice of the inverse auto-regressive flow, sampling
is possible via one forward pass, but evaluating the density is done sequentially across each component.
Doubling the resolution, i.e. using $\nside = 64$ takes $\sim 1.2$s and $\nside = 128$
takes $\sim 2.4$s.

\begin{figure}
    \centering
    \includegraphics[width=\textwidth]{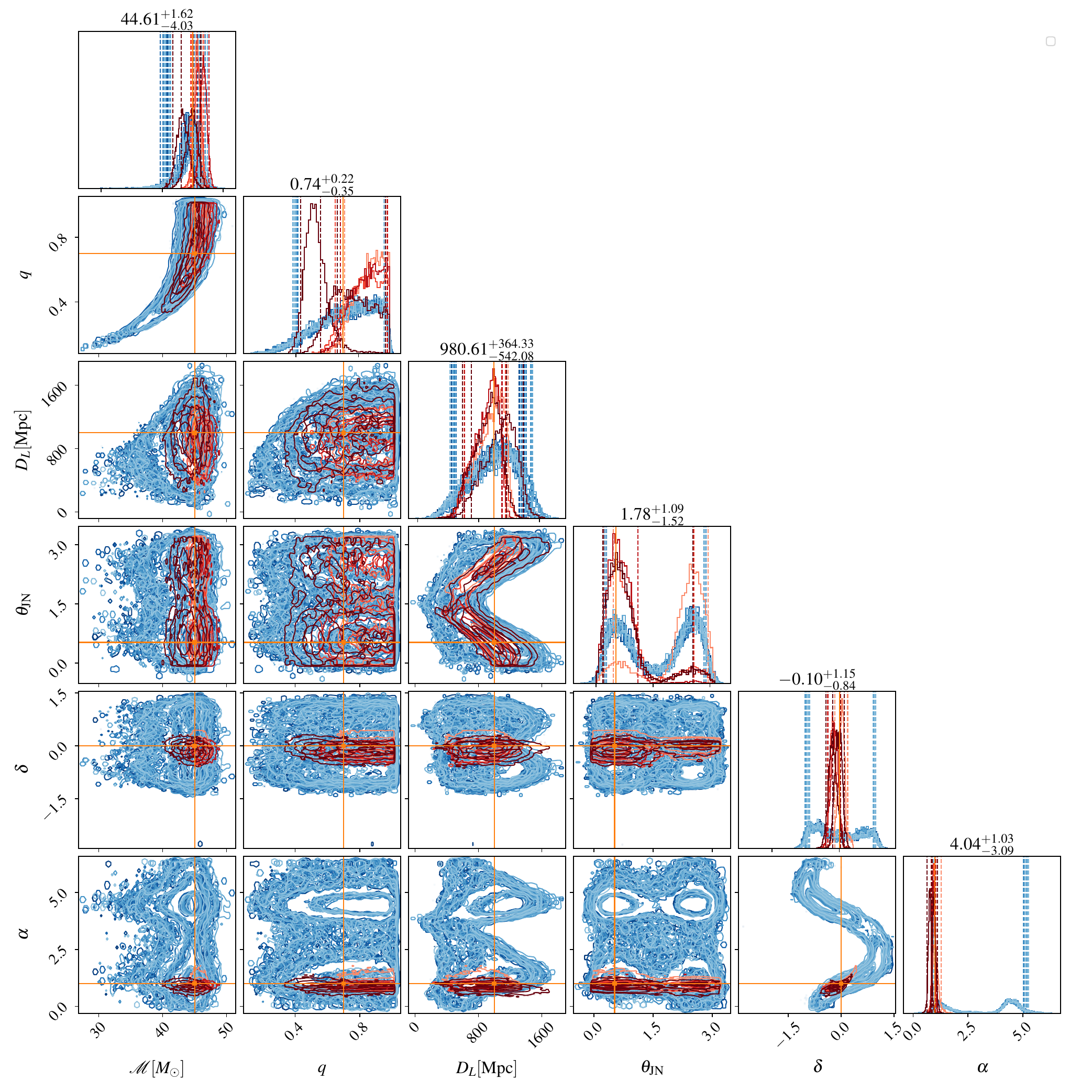}
    \caption{Example posterior for a signal with parameters $\{\mc=45\msun, q = 0.7, \dl = 1000~\mpc\}$ injected in
    20 different background instances, sampled using {\amplfi} is shown in sky blue . All posteriors are consistent
    with one another. Posteriors from the same signal injected in 5 different background instances (same background stretch
    as the {\amplfi} injections) and analyzed via nested sampling with Bilby, is overlayed in varying Orange-red colors.
    We find that parameters like {\mc} and $q$ are consistent in terms of detection uncertainties across different
    runs. Extrinsic parameters, especially the sky-location, though consistent with the true parameters, shows larger
    uncertainty with {\amplfi}.
    }
    \label{fig:high-mass-compare}
\end{figure}
\begin{figure}
    \centering
    \includegraphics[width=\textwidth]{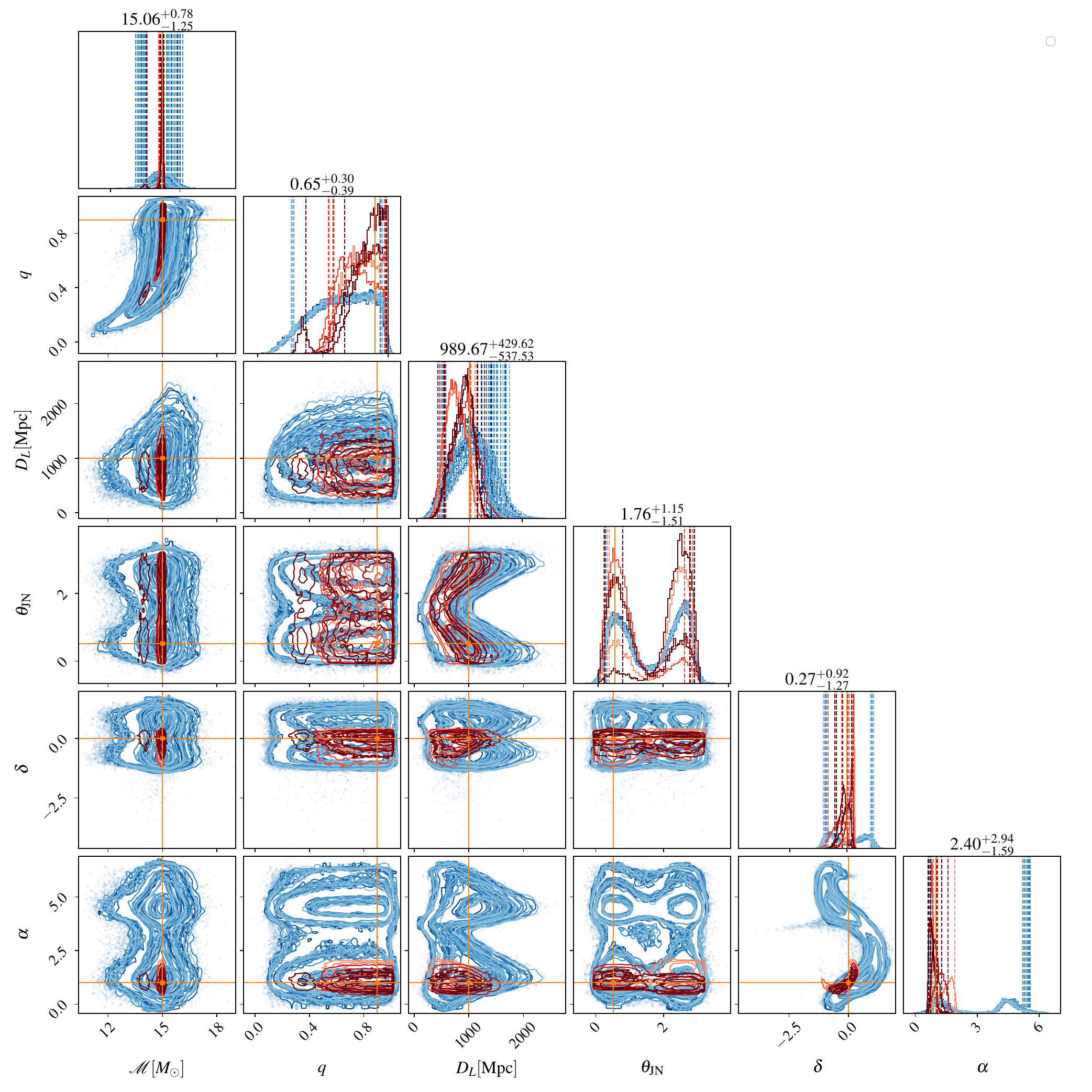}
    \caption{Example posterior for a signal with parameters $\{\mc=15\msun, q = 0.9, \dl = 1000~\mpc\}$ injected in
    20 different background instances, sampled using {\amplfi} is shown in sky blue. All posteriors are consistent
    with one another. Posteriors from the same signal injected in 5 different background instances (same background stretch
    as the {\amplfi} injections) and analyzed via nested sampling with Bilby, is overlayed in varied Orange-red colors.
    We find that parameters like {\mc} and $q$ are consistent in terms of detection uncertainties across different
    runs. Extrinsic parameters, especially the sky-location, though consistent with the true parameters, shows larger
    uncertainty with {\amplfi}.
    }
    \label{fig:low-mass-compare}
\end{figure}
\begin{figure}
    \centering
    \includegraphics[width=0.6\textwidth]{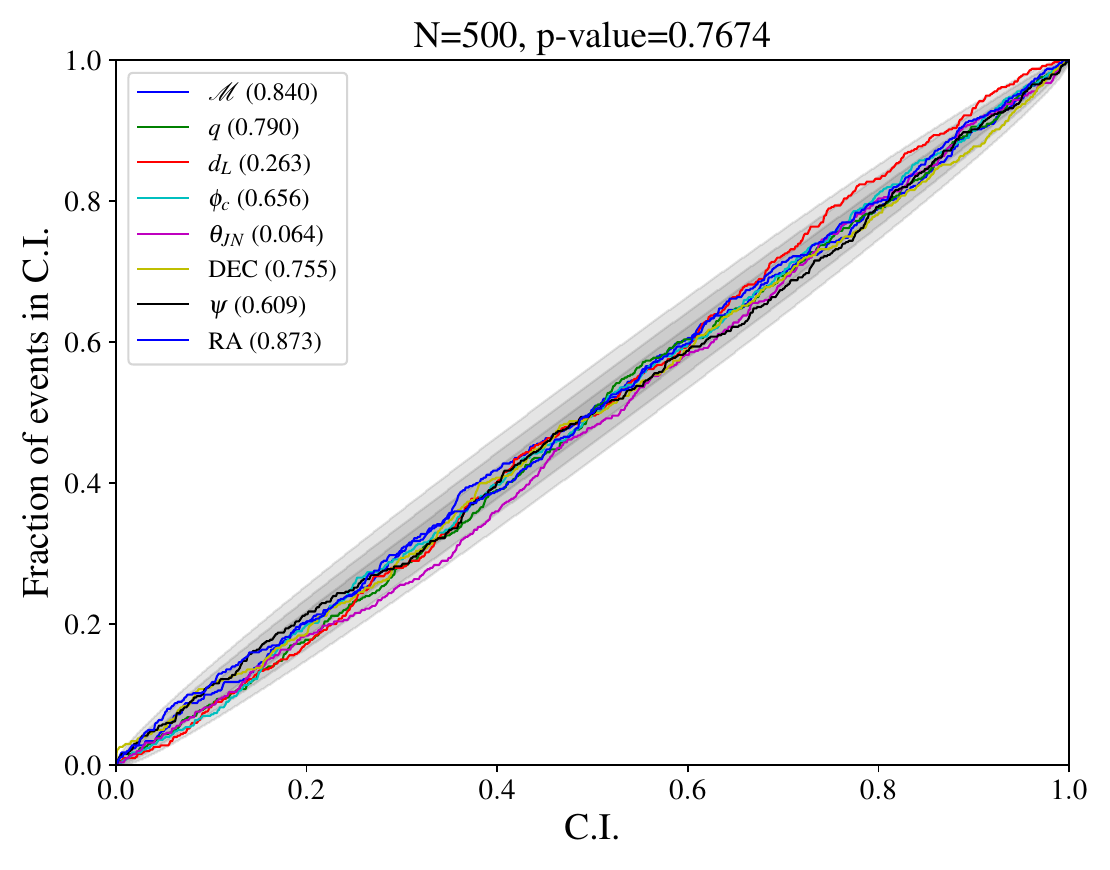}
    \includegraphics[width=0.38\textwidth, trim=0.3cm 0.3cm 0.3cm 0.3cm, clip]{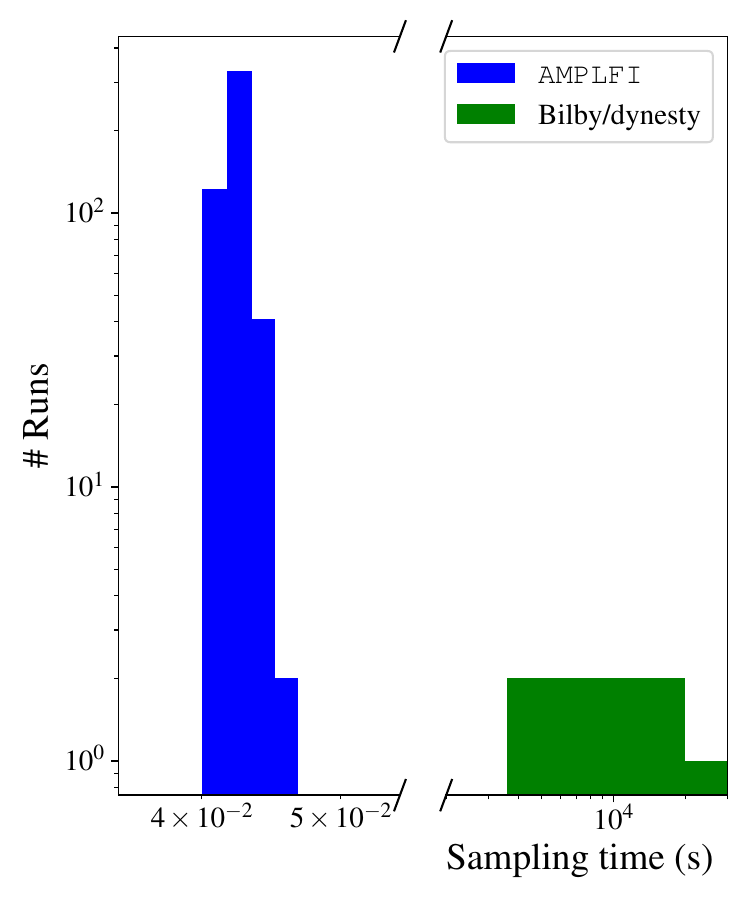}
    \caption{\textbf{Left}: Percentile-percentile (PP) plot showing recovery accuracy for 500
    BBH injections performed in a testing background, different from training background.
    The different lines track the cumulative fraction of events within a corresponding
    confidence interval for the parameters mentioned in Eq.~\ref{eq:parameters}.
    \textbf{Right}: Sampling times for {\amplfi} vs. nested sampling
    runs done on injections using Bilby, with identical waveform model and prior settings.
    The nested sampling runs were done with a CPU pool size of 24, and correspond to the runs
    using in Figure~\ref{fig:low-mass-compare}. The standard GW likelihood model is used.
    The {\amplfi} sampling times correspond to the 500 injections used for the P-P plot on the left.}
    \label{fig:pp-plot}
\end{figure}
\begin{figure}
    \centering
        \includegraphics[width=0.8\textwidth]{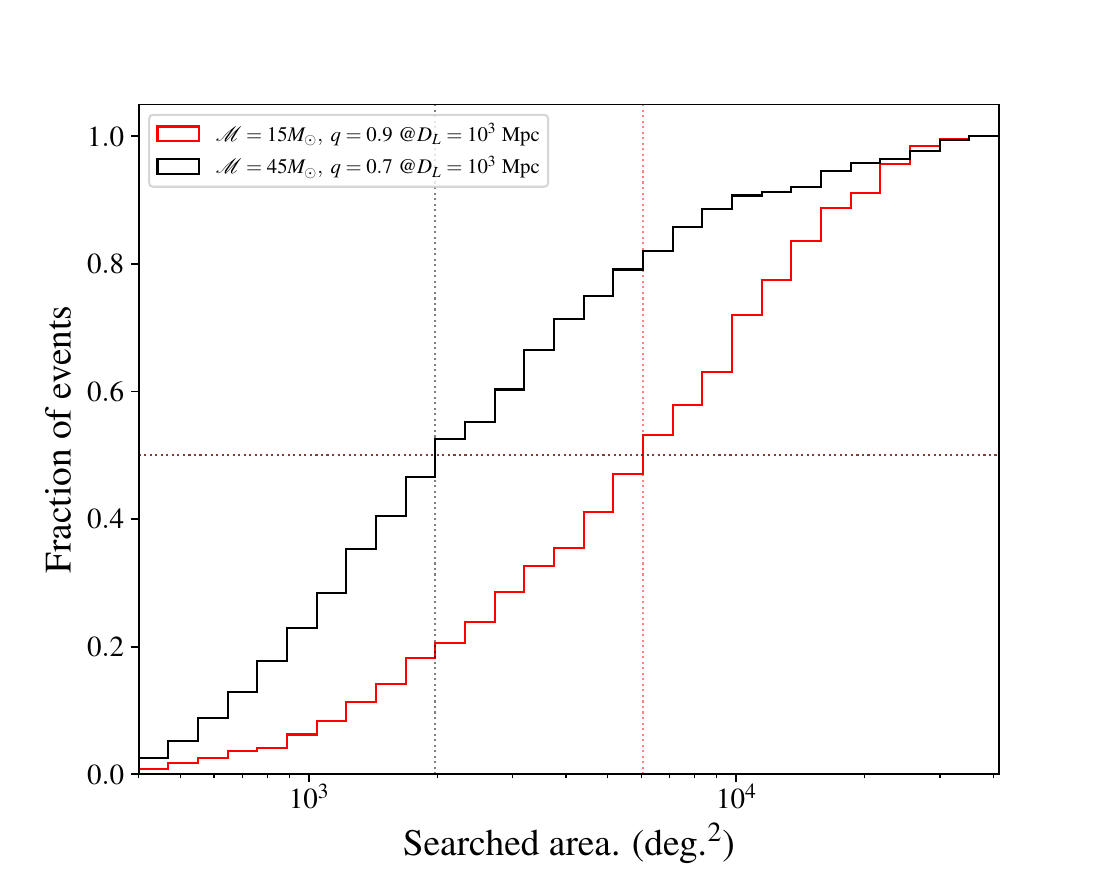}
    \caption{Searched area for a representative system with $\mc = 45\msun$ and
    $\mc = 15\msun$ at $\dl = 1$ Gpc done across the sky. We note that, as expected,
    larger chirp mass give better searched area as a result of being louder signals. However,
    the median searched area is $\mathcal{O}(1000) \mathrm{deg}^2$.}
    \label{fig:searched-area}
\end{figure}
\section{\label{sec:results}Results and Performance}
We show example posteriors in Figures~\ref{fig:high-mass-compare} and \ref{fig:low-mass-compare} from representative
higher and lower mass BBH signals. In both cases, the signal injection is performed in a background different from that
used during training. The same signal is injected 20 times at different background segments and samples
are drawn. The posteriors are shown in the blue colormap. We also perform inference on the same system via nested sampling
using \texttt{Bilby}~\cite{Ashton:2018jfp,Romero_Shaw_2020} and \texttt{Dynesty}~\cite{Speagle:2019ivv} with 1500 live points and identical
priors as {\amplfi}, repeating 5-times on different background segments.
This is shown in the red colormap. We see that there is good consistency among the {\amplfi} posteriors i.e., distributions in blue colormap
fall on top of each other. For higher masses, like the $\mc = 45\;\msun$, shown in Figure~\ref{fig:high-mass-compare},
we find the recovery accuracy of the chirp mass to be comparable to nested sampling. Also, considering all nested sampling
runs, the mass-ratio accuracy from {\amplfi} is comparable. However, extrinsic parameters like the
inclination, or the sky location are not recovered with the same accuracy as nested sampling, although broad features
like the ``ring'' pattern in the skymap are evident. The same is also true for the inclination posterior, where the
inference is degenerate between the true inclination angle, and its supplementary angle. In case of the nested sampling
runs, though this degeneracy is broken, the inference may not always select the right ``peak'' for the inclination.
Thus, overall, we find consistency of the {\amplfi} results with the true parameters of the injection and with nested
sampling results, though not as accurate for some parameters.

We also find that the inference for lower mass, $\mc=15\;\msun$ BBH system, shown in Figure~\ref{fig:low-mass-compare},
is worse compared to higher mass. For example, in Figure~\ref{fig:low-mass-compare}, we see that the $\mc$ posteriors
recovered by {\amplfi} is broader compared to that recovered from nested sampling. The extrinsic parameter recovery
follows similar pattern as the high-mass example. Though broad features of the sky location like the ``ring'' pattern
for two detector is evident, the recovery is not at the level of nested sampling results. The greater consistency of the
intrinsic parameters compared to the extrinsic parameters suggests one potential avenue of improvement being to further
augment our dataloader in the extrinsic parameters $(\alpha, \delta, \psi)$ which is used to project the signal onto
the GW antennae. Signal projection as implemented in {\mlgw} performs this operation on-the-fly on the GPU,
hence oversampling the corresponding priors and creating a larger batch is feasible without compromising data
generation time, and will be considered in a future implementation.

In Figure~\ref{fig:pp-plot}, we show parameter recovery consistency with true values via a percentile-percentile (PP)
plot for 500 simulated BBHs. The parameters of these are sampled from the same prior as that used during training.
As for the previous examples above, for these injections, the background is different from the training background. The diagonal
trend of the plot shows that there is no bias in the inference with {\amplfi} across the parameter space. In Figure~\ref{fig:searched-area},
we show the searched area distributions for the two representative higher mass BBH and lower mass BBH placed at a fiducial
distance of $\dl = 1000\;\mpc$. We see the expected trend of getting a better search area statistic for higher mass
system compared to a lower mass system due to their larger amplitude. However, the searched area is several
$\mathcal{O}(1000)$ degrees for the two detector model because of the larger error-bars on the sky coordinates
mentioned above.

\subsection{Comparison with BAYESTAR skymaps}
\begin{figure}[t]
    \centering
    \scriptsize{$\mathrm{S190512at}$} \\
    \includegraphics[width=0.485\textwidth]{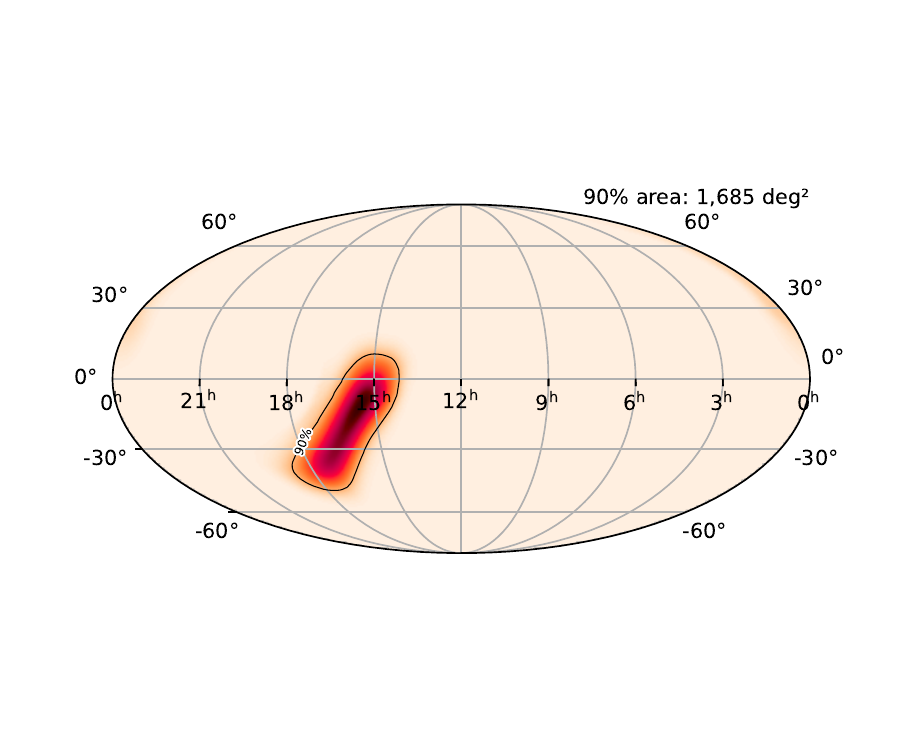} \includegraphics[width=0.485\textwidth]{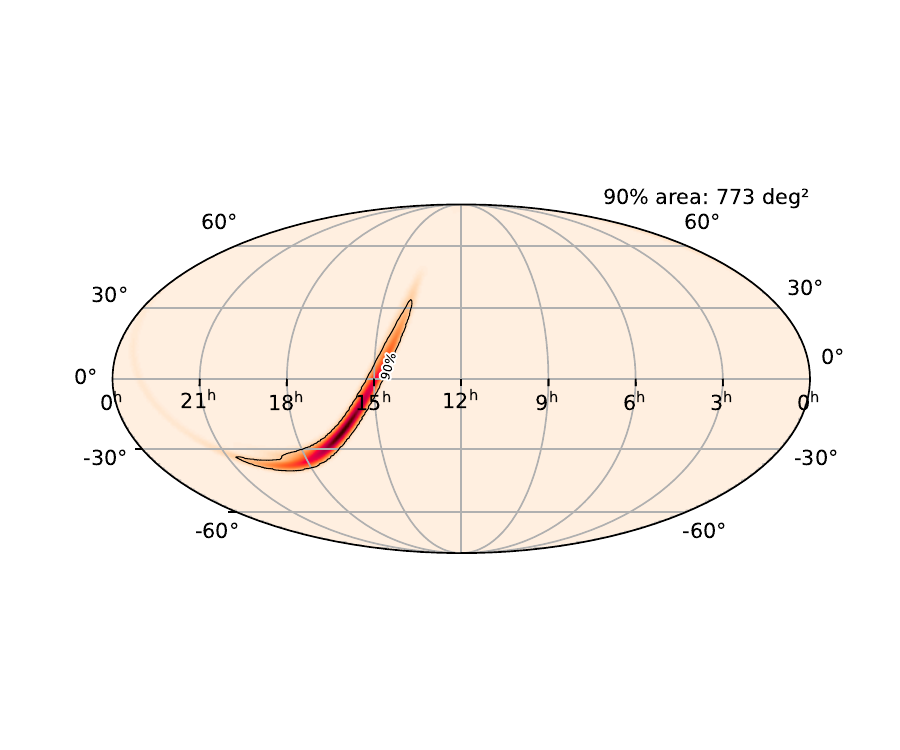}\\
    \scriptsize{$\mathrm{S190513bm}$} \\
    \includegraphics[width=0.485\textwidth]{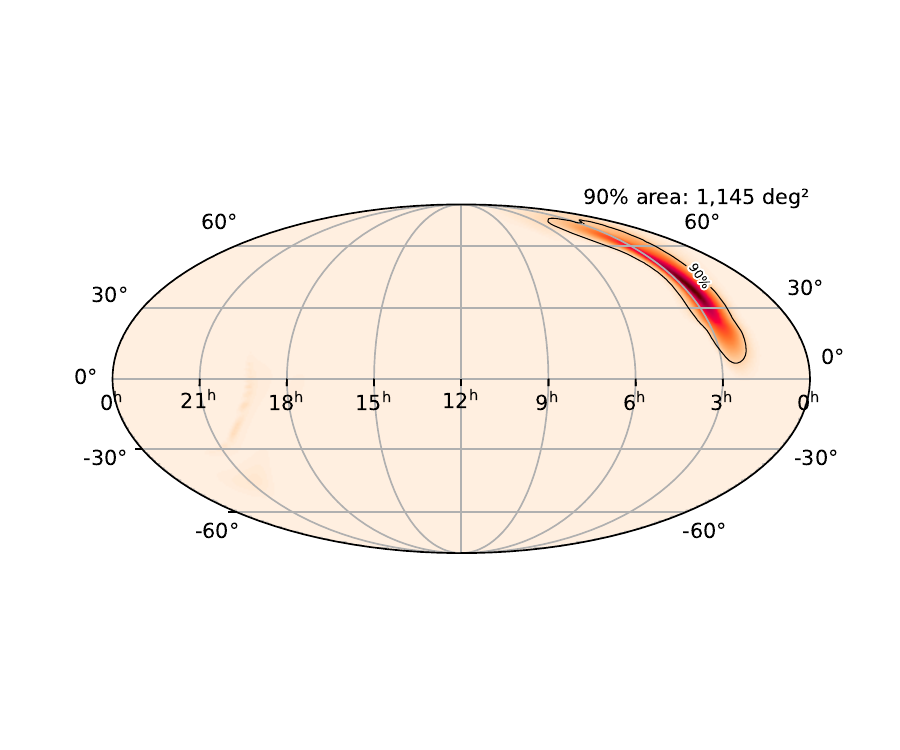} \includegraphics[width=0.485\textwidth]{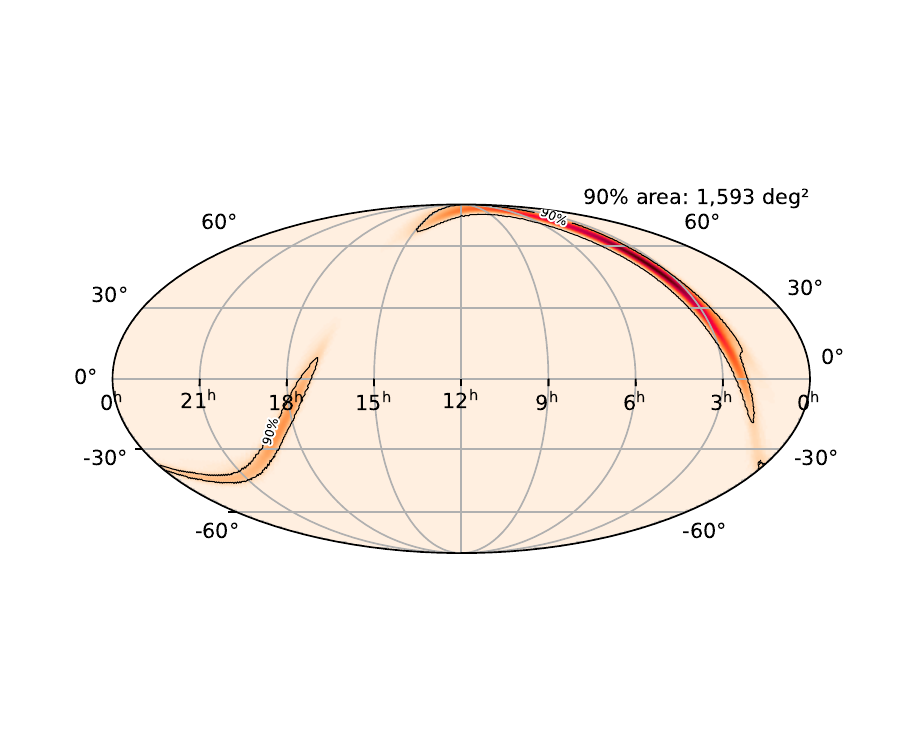}\\
    \scriptsize{$\mathrm{S190701ah}$} \\
    \includegraphics[width=0.485\textwidth]{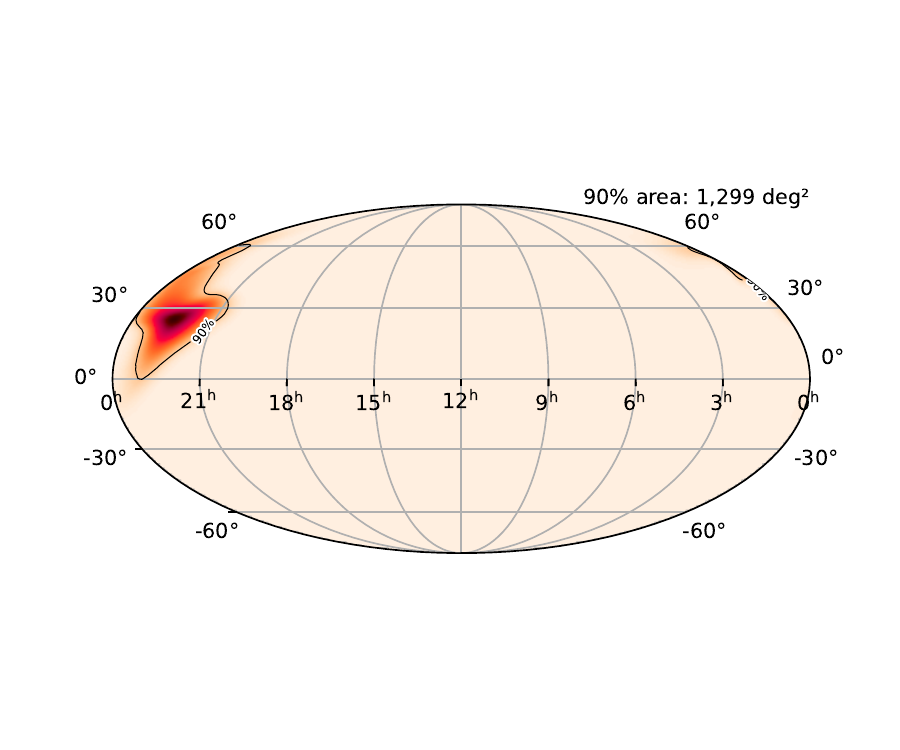} \includegraphics[width=0.485\textwidth]{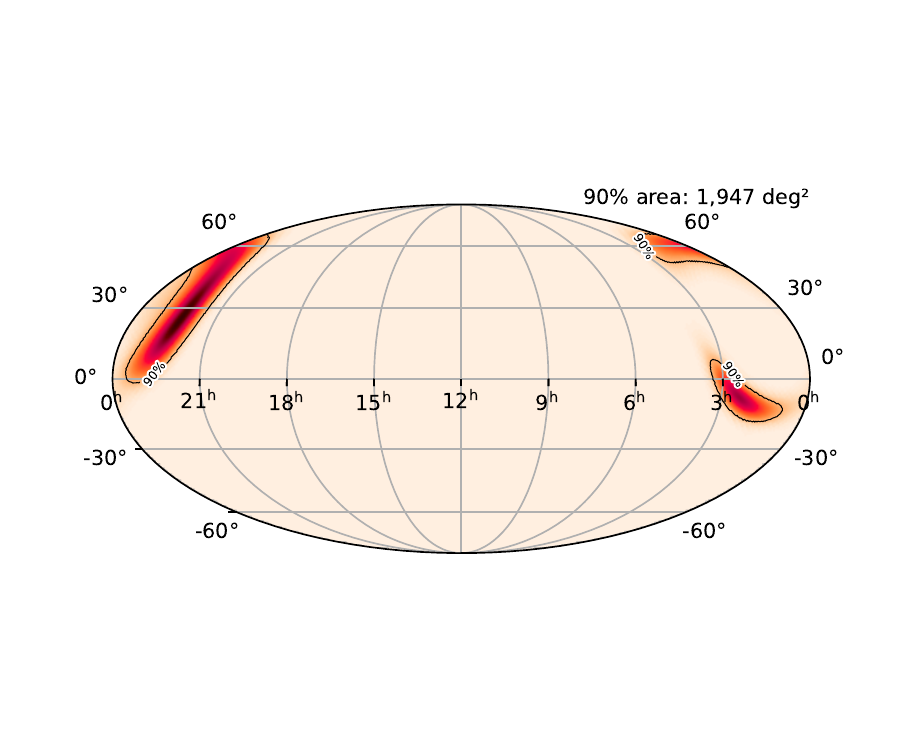}\\
    \caption{\textbf{Left}: {\amplfi} skymaps \textbf{Right} Bayestar maps from events, using HL data, spanning different
    months of the third observing run, O3. We find that the skymaps are broadly consistent, although there is no clear trend
    of one being more constraining. However, it demonstrates model validity to differing backgrounds up to several months from that
    used in training.}
    \label{fig:bayestar-maps}
\end{figure}
\begin{figure}[t]
   \ContinuedFloat
    \centering
    \scriptsize{$\mathrm{S191215w}$} \\
    \includegraphics[width=0.485\textwidth]{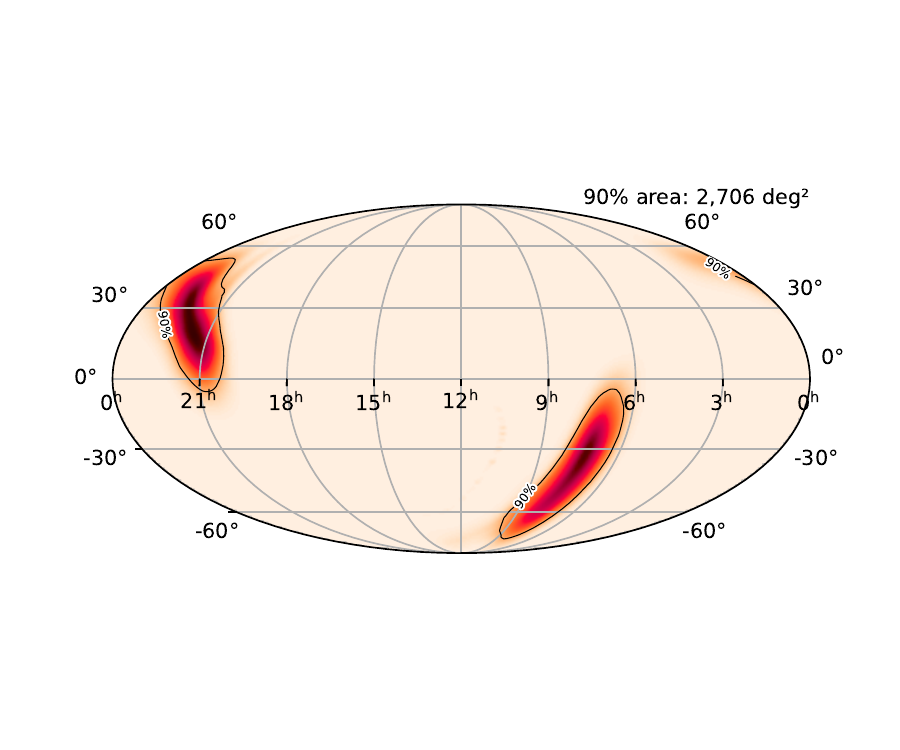} \includegraphics[width=0.485\textwidth]{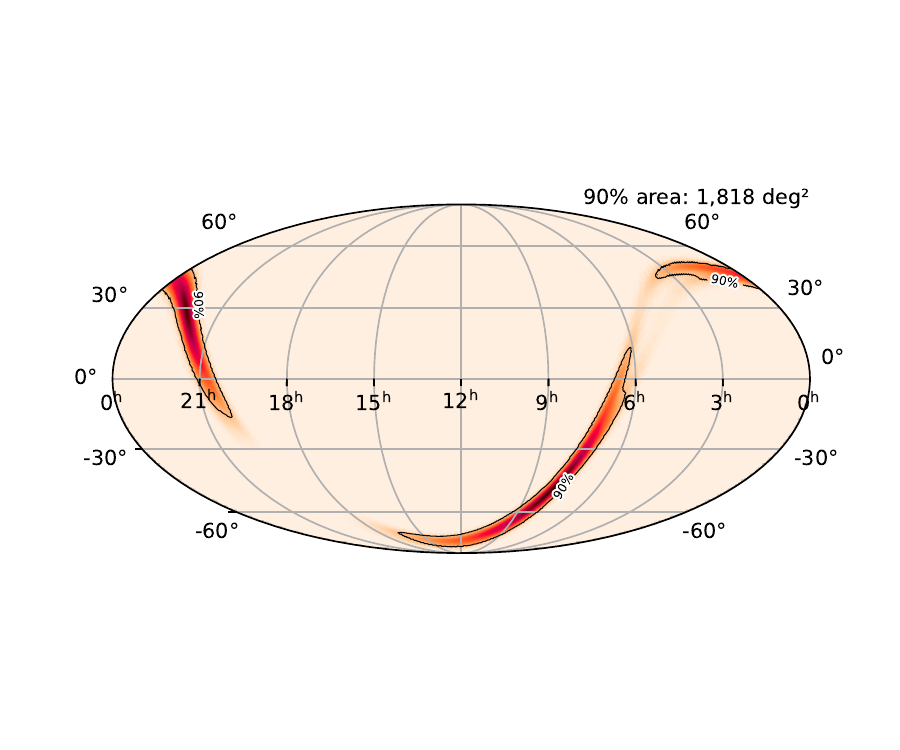}\\
    \scriptsize{$\mathrm{S200129m}$} \\
    \includegraphics[width=0.485\textwidth]{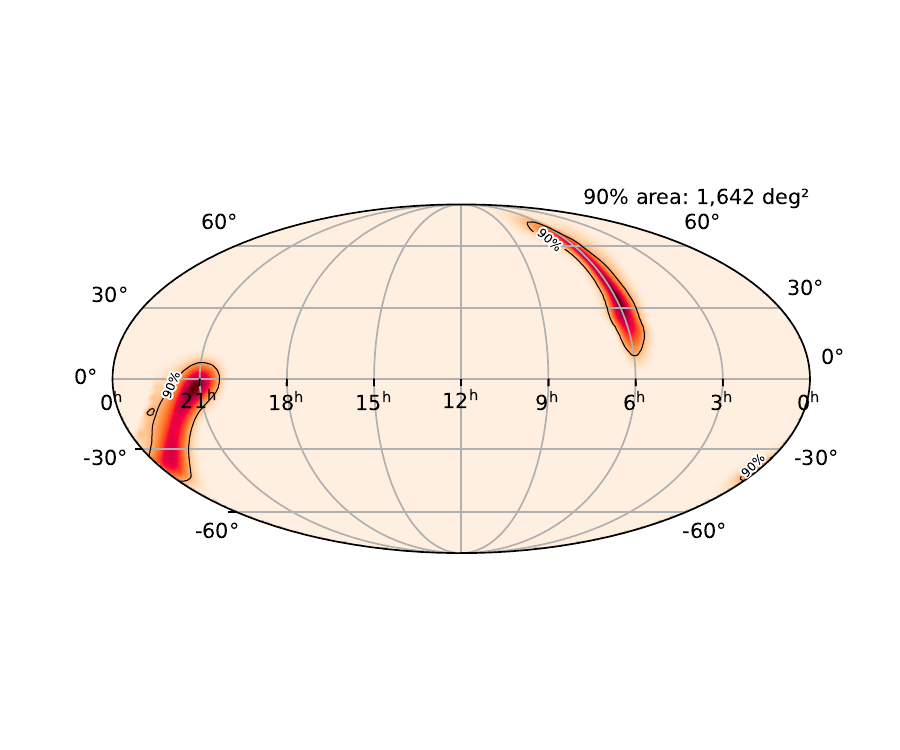} \includegraphics[width=0.485\textwidth]{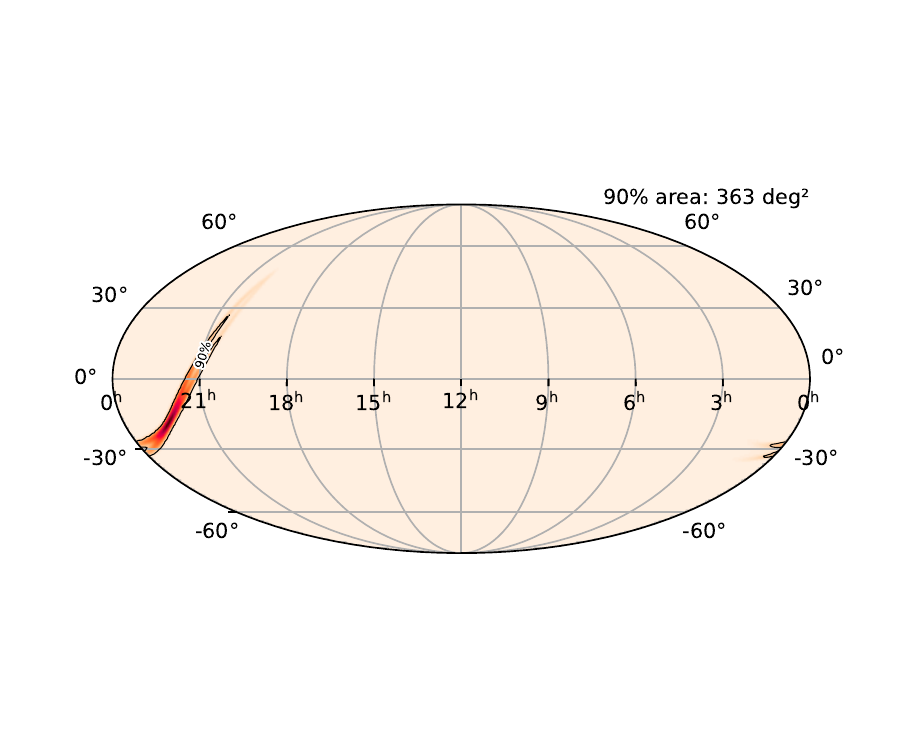}\\
    \scriptsize{$\mathrm{S200311bg}$} \\
    \includegraphics[width=0.485\textwidth]{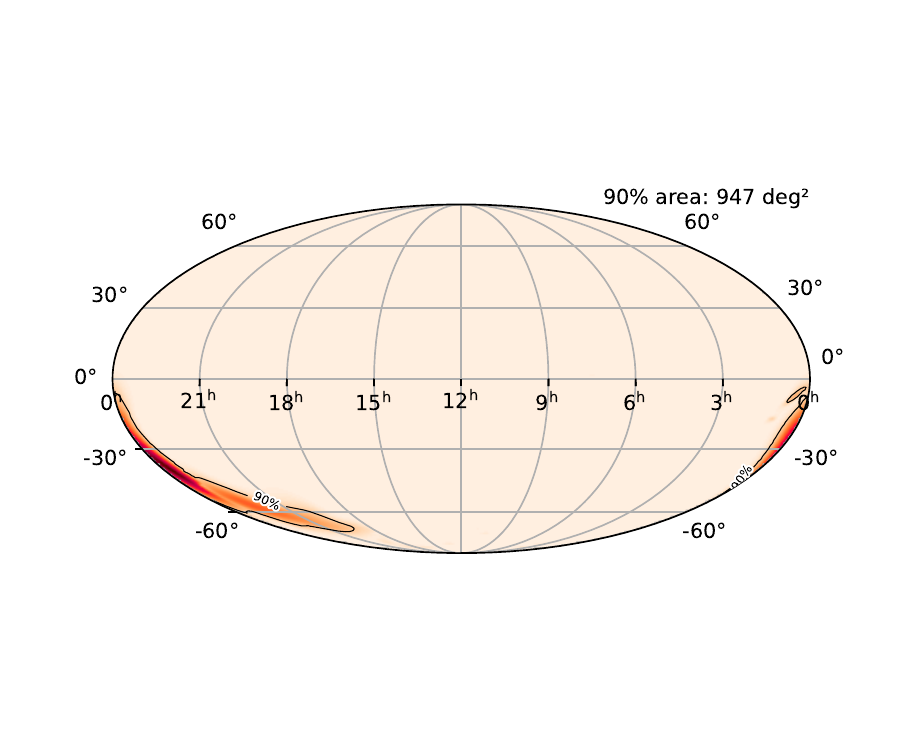} \includegraphics[width=0.485\textwidth]{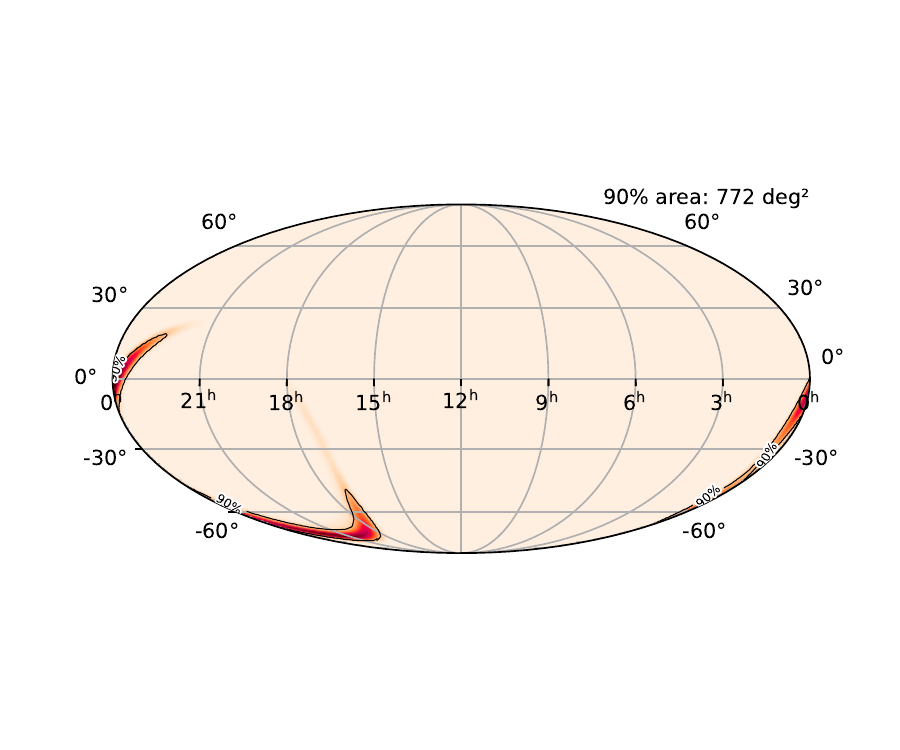}\\
    \caption{Continued from above -- \textbf{Left}: {\amplfi} skymaps \textbf{Right} Bayestar maps.}
  \label{fig:results}
\end{figure}
In this section, we run our model on several O3 events using the data from GWOSC and compare the sky-localization with the
corresponding rapid localization method, BAYESTAR~\cite{Singer_2016}. Since our model is trained on only 2-detector (HL)
data, we re-run BAYESTAR on the events using only HL SNR timeseries for this comparison.
We show the skymap, along with the 90\% localization
area, for several O3 events that were detected online in Figure~\ref{fig:bayestar-maps}. The left (right) panel shows the
reconstruction using {\amplfi} (BAYESTAR). The choice of the selected events is that their
event parameters, as published in GWTC-3, lie within the prior range used by us during training, and
the events are spread over the most of O3 from May 2019 to March 2020 (the event identifiers carry
the date of discovery). While our training background is a half-day chunk at the start of O3,
we test on events which had been detected over the duration of the entire run. This is done to obtain
a qualitative assessment of the impact of changing background i.e., if the model performance greatly degrades when
supplied with data several months away from the training background. We find that there is broad similarity in the
skymaps between {\amplfi} and BAYESTAR. There is no trend in terms of the skymaps being more/less constrained.
However, given that the events used for Figure~\ref{fig:bayestar-maps} range from May 2019 to March 2020, (recall that
training background is limited to half a day in May 2019), we conclude that the model validity is maintained for
different background up to several months. This does not imply that the model once trained, may be optimal for the
entire run. In fact, one of the requirements for {\amplfi} is the ability for periodic re-training; however,
we anticipate that such re-training may converge quickly given the preliminary observations mentioned here.
Furthermore, the suitable re-training cadence is yet to be determined and will be reported in the future.

\section{Conclusion and Future Work}\label{sec:conclusion}
In this work, we presented a GW parameter estimation algorithm, {\amplfi}, using likelihood-free inference.
This work is one of the efforts to integrate AI algorithms in GW data analysis using tools build as a part of
{\mlgw} -- like data cleaning using {\tt{DeepClean}}~\cite{saleem2023demonstration}, search of CBCs using
{\aframe}~\cite{aframe},
anomaly detection algorithm {\tt GWAK}~\cite{Raikman:2023ktu}.
The use case of {\amplfi} is to run alongside recently reported neural-network based CBC search, {\aframe}.
The intended design is for both {\aframe~+~\amplfi} to run online, preferably on the same hardware to minimize
any communication overhead and reduce the time-to-alert. 
While to date we have focused on this integration of neural-networks in order to achieve very low-latency outputs, there is reason in principle that {\amplfi} could not also be paired with any other search pipeline that provides an estimated time of arrival.
An important future product of this research will be a standalone version of the software that can be run simply with the provision of trigger time, enabling easy adoption for searches that want to use it.
Current real-time GW alert data products include
sky-localization and EM-bright source properties apart from the significance, and a derived data-product,
P-astro, based on rate of foreground and background triggers. We have not discussed the EM Bright source
properties in this work since the analysis, so far, is limited to BBH signal which are expected to be
EM-dark. However, the availability of the posterior samples in real-time allows for their straightforward
computation by binning the posterior samples, or marginalizing them over several equations of state (see
section C.2 of \cite{Chaudhary_2024}).

As a part of routine end-to-end testing, the LVK has set up a streaming data channel with injections over
a 40-days chunk of O3 background. This was used to profile latencies in several components in the alert
infrastructure, reported in \cite{Chaudhary_2024}. Preliminary work has been done
toward the online deployment of {\aframe~+ \amplfi} to analyze this data stream. Based on preliminary
testing, we find the net latency of data acquisition by \aframe, evaluating significane, passing data to
\amplfi, followed by generating posteriors and skymaps takes $\sim 6$ seconds. At the time of writing,
candidates are reported to a test instance of GraceDB, the candidate database used by the LVK, however,
the view for the same is not public. As a follow-up to the methods reported here, the performance
of {\aframe~+ \amplfi} will be reported on this mock data challenge (MDC) in a future work. The dataset
contains $\mathcal{O}(1000)$ BBHs, for which several match-filtering searches and annotation pipelines including
BAYESTAR was run. This will contain a systematic comparison with BAYESTAR on the simulated BBHs in the MDC,
along with comparison with online PE results reported as a part of the same study.

Certain aspects of the model requires improvements, for example, the inference on the extrinsic parameters
like sky location, and the extension to use spinning waveforms. This will be considered in a future work.
Also, we note that the focus has been on BBHs due to their shorter signal duration. However, the main focus
of MMA is binary neutron star (BNS) and neutron star black hole (NSBH) systems for which the signal duration
can be $\mathcal{O}$(min) depending on the starting frequency. This makes the
input arrays larger by an order of magnitude compared to the ones considered here, and therefore expensive
in terms of memory and compute. However, low mass systems \emph{inspiral} for most of that duration and the
frequency evolution is described analytically, primarily at newtonian order. Thus, feature extraction
from the time-series data, or alternative data representations like q-transforms can be a possible
approach toward search and parameter estimation.

Finally, the framework for {\amplfi} is not limited to CBC signals, and can be extended to burst
signal morphologies like sine-Gaussians. This is relevant
for running parameter estimation on candidates picked up by pipelines like
{\tt GWAK} that look for unmodeled events~\cite{Raikman:2023ktu}. In this case, a sine-gaussian parameter
estimation may lead to measurement of fundamental features like central frequency or duration.

\section{Acknowledgments}
Authors acknowledge support from NSF PHY-2117997 ``Accelerated AI Algorithms for Data-Driven Discovery (A3D3).''
This work used NCSA-Delta at U. Illinois through allocation PHY-240078 from the Advanced Cyberinfrastructure
Coordination Ecosystem: Services \& Support (ACCESS) program, which is supported by National Science Foundation
grants \#2138259, \#2138286, \#2138307, \#2137603, and \#2138296. This work made use of resources provided by
subMIT at MIT Physics.
This research has made use of data or software obtained from the Gravitational Wave Open Science Center
(gwosc.org),\cite{gwosc1, gwosc2} a service of the LIGO Scientific Collaboration, the Virgo Collaboration,
and KAGRA. This material
is based upon work supported by NSF's LIGO Laboratory which is a major facility fully funded by the NSF, as
well as the Science and Technology Facilities Council (STFC) of the United Kingdom, the Max-Planck-Society
(MPS), and the State of Niedersachsen/Germany for support of the construction of Advanced LIGO and construction
and operation of the GEO600 detector. Additional support for Advanced LIGO was provided by the Australian
Research Council. Virgo is funded, through the European Gravitational Observatory (EGO), by the French Centre
National de Recherche Scientifique (CNRS), the Italian Instituto Nazionale di Fisica Nucleare (INFN) and the
Dutch Nikhef, with contributions by institutions from Belgium, Germany, Greece, Hungary, Ireland, Japan,
Monaco, Poland, Portugal, Spain. KAGRA is supported by Ministry of Education, Culture, Sports, Science and
Technology (MEXT), Japan Society for the Promotion of Science (JSPS) in Japan; National Research Foundation
(NRF) and Ministry of Science and ICT (MSIT) in Korea; Academia Sinica (AS) and National Science and
Technology Council (NSTC) in Taiwan.

\clearpage
\appendix
\section{\label{appendix:embedding-hpo}Hyperparameter Tuning of the Embedding Network}
\begin{table}
\caption{Top 10 hyper-parameter configurations for hyper-parameter optimization runs for the embedding
network. The best trial configuration is shown in boldface.}
\footnotesize
\begin{tabular}{ccccccccccc}
\br
ResNet conf. & kernel size & $\lambda_1$ & $\lambda_2$ & $\lambda_3$ & $D_{\gamma}$ & LR & Momentum & wt. decay & $N$ & VICReg. \\
\br
( $\bf{5}$ , $\bf{3}$ , $\bf{3}$ ) &  $\bf{5}$  &  $\bf{1}$  &  $\bf{1}$  &  $\bf{5}$  &  $\bf{8}$  &  $\bf{7.16\cdot 10^{-4}}$  &  $\bf{8.07\cdot 10^{-5}}$  &  $\bf{4.42\cdot 10^{-4}}$  &  $\bf{3}$  &  $\bf{0.48}$ \\
( $4$ , $3$ , $3$ ) &  $7$  &  $1$  &  $1$  &  $1$  &  $11$  &  $8.97\cdot 10^{-4}$  &  $5.75\cdot 10^{-5}$  &  $9.01\cdot 10^{-5}$  &  $3$  &  $0.528$ \\
( $4$ , $5$ , $3$ ) &  $7$  &  $1$  &  $1$  &  $1$  &  $9$  &  $2.08\cdot 10^{-4}$  &  $2.60\cdot 10^{-4}$  &  $3.72\cdot 10^{-4}$  &  $3$  &  $0.539$ \\
( $4$ , $3$ , $3$ ) &  $7$  &  $1$  &  $1$  &  $1$  &  $10$  &  $2.62\cdot 10^{-4}$  &  $3.69\cdot 10^{-5}$  &  $1.16\cdot 10^{-5}$  &  $3$  &  $0.543$ \\
( $4$ , $5$ , $3$ ) &  $5$  &  $5$  &  $1$  &  $1$  &  $8$  &  $3.48\cdot 10^{-4}$  &  $6.10\cdot 10^{-4}$  &  $7.37\cdot 10^{-4}$  &  $5$  &  $0.553$ \\
( $5$ , $5$ , $4$ ) &  $3$  &  $5$  &  $1$  &  $1$  &  $8$  &  $1.31\cdot 10^{-4}$  &  $1.75\cdot 10^{-5}$  &  $1.67\cdot 10^{-4}$  &  $3$  &  $0.578$ \\
( $3$ , $4$ , $4$ ) &  $3$  &  $5$  &  $1$  &  $5$  &  $9$  &  $5.38\cdot 10^{-4}$  &  $1.42\cdot 10^{-5}$  &  $1.67\cdot 10^{-5}$  &  $5$  &  $0.610$ \\
( $4$ , $5$ , $3$ ) &  $3$  &  $1$  &  $1$  &  $5$  &  $10$  &  $9.76\cdot 10^{-4}$  &  $2.69\cdot 10^{-4}$  &  $4.20\cdot 10^{-5}$  &  $4$  &  $0.627$ \\
( $4$ , $3$ , $4$ ) &  $3$  &  $1$  &  $1$  &  $1$  &  $9$  &  $1.37\cdot 10^{-4}$  &  $5.46\cdot 10^{-5}$  &  $1.33\cdot 10^{-5}$  &  $3$  &  $0.677$ \\
( $4$ , $5$ , $3$ ) &  $5$  &  $5$  &  $5$  &  $1$  &  $8$  &  $1.64\cdot 10^{-4}$  &  $9.47\cdot 10^{-4}$  &  $1.61\cdot 10^{-5}$  &  $3$  &  $0.707$ \\
\hline
\end{tabular}
\label{tab:embedding-hpo-table}
\end{table}
\begin{figure}[h]
    \centering
    \includegraphics[width=0.7\textwidth]{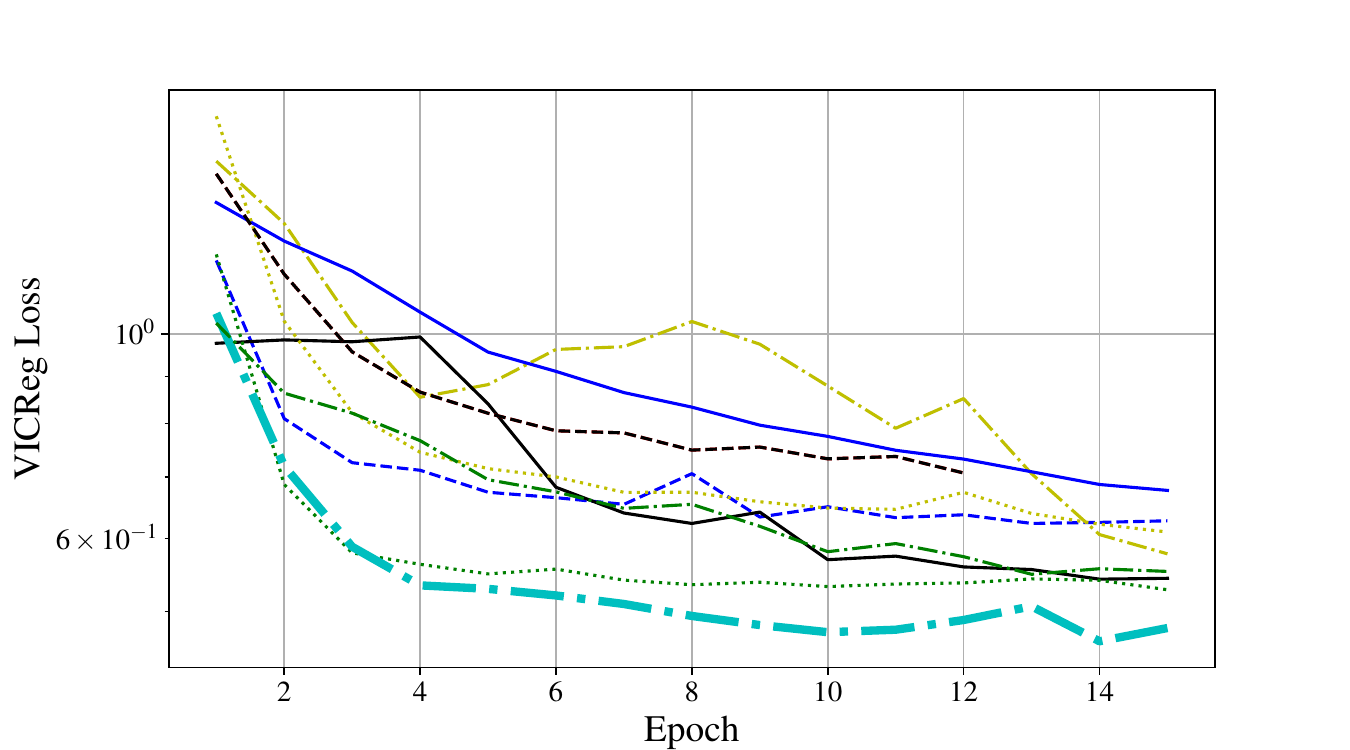}
    \includegraphics[width=0.28\textwidth]{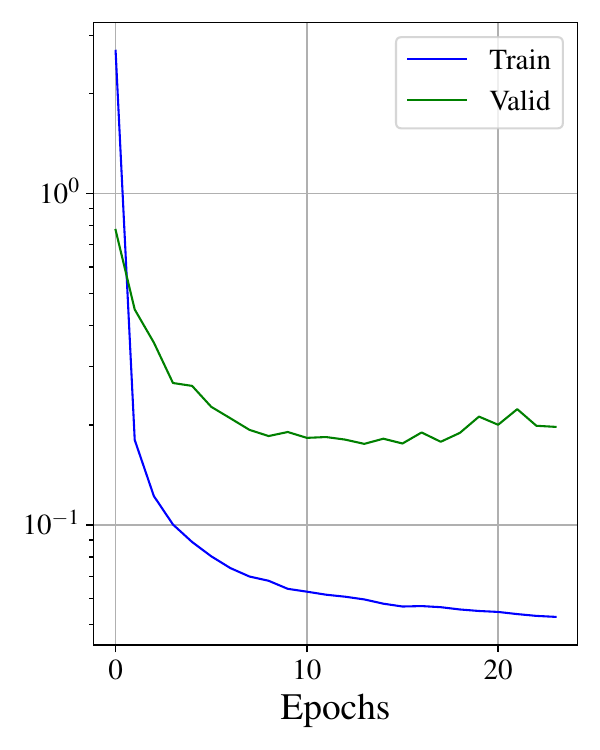}
    \caption{\textbf{Left}: Avg. VICReg validation loss as a function of training epoch from the
    top 10 HPO runs in Table~\ref{tab:embedding-hpo-table}. The best configuration is plotted in thick
    solid line; corresponding configuration is boldfaced entry in Table~\ref{tab:embedding-hpo-table}.
    \textbf{Right}: Training/Validation loss for best model configuration from the left panel until
    early-stopping.}
    \label{fig:embedding-hpo-top-runs}
\end{figure}
To determine the configuration to be used for the embedding network mentioned in Sec.~\ref{sec:embedding-net},
we perform an extensive hyperparameter optimization over the search space of the layers of ResNet, the
convolutional kernel size of the ResNet, the dimensionality of the representation, $D_{\gamma}$, the
dimensionality of the expanded space where the VICReg loss is computed; this is tune by a factor, $N$,
i.e. the dimensionality of the expanded representation is $N\times D_{\gamma}$. We use Stochastic gradient
descent optimizer and also sample over the learning rate, the weight decay and the momentum terms of the
optimizer. A total of $\sim 250$ training runs were performed using asynchronous hyperbanding with
early-stopping~\cite{li2018massively} using the {\raytune} library~\cite{liaw2018tune}. This
technique stops poor performing trials allowing more favorable trials to continue. We use a
grace period of 3 epochs before half of ongoing trials are stopped. The experiment was carried over
8 workers over 4 A40 GPU taking $\sim 40$ hours. We show the top 10 trial
configuration in Table~\ref{tab:embedding-hpo-table} and show the epoch-average VICReg validation
loss for the same in the left panel of Figure~\ref{fig:embedding-hpo-top-runs}. 
The right panel of the figure shows the training/validation of the best model configuration
trained until early-stopping condition is met.

\section{\label{appendix:flow-hpo}Hyperparameter Tuning of the Normalizing Flow}
\begin{table}
    \caption{Top 10 hyper-parameter configurations for run involving 30 epochs. Validation loss
    is noted at the end of the 30th epoch.}
    \footnotesize
    \begin{tabular}{ccccccc}
    \br
    \# transforms & \# blocks & \# hidden feat. & LR & batch size & wt. decay & val. loss
    \\ \hline
    $\bf{60}$ & $\bf{6}$ & $\bf{100}$ & $\bf{0.00129}$ & $\bf{800}$ & $\bf{0.00241}$ & $\bf{8.52}$ \\
    $100$ & $6$ & $150$ & $0.000636$ & $1000$ & $0.000263$ & $8.63$ \\
    $60$ & $6$ & $120$ & $0.00242$ & $800$ & $0.00471$ & $8.65$ \\
    $60$ & $8$ & $150$ & $0.000488$ & $1000$ & $0.000553$ & $8.70$ \\
    $80$ & $6$ & $120$ & $0.000442$ & $1000$ & $0.00161$ & $8.71$ \\
    $60$ & $7$ & $150$ & $0.00103$ & $1000$ & $0.000271$ & $8.72$ \\
    $80$ & $7$ & $120$ & $0.000318$ & $800$ & $0.000173$ & $8.78$ \\
    $80$ & $6$ & $100$ & $0.00093$ & $1000$ & $0.00332$ & $8.83$ \\
    $80$ & $8$ & $100$ & $0.000873$ & $1200$ & $0.00453$ & $8.86$ \\
    $60$ & $6$ & $120$ & $0.000297$ & $1000$ & $0.0509$ & $8.87$ \\
    \hline
    \end{tabular}
    \label{tab:flow-hpo-table}
\end{table}
\begin{figure}[h]
    \centering
    \includegraphics[width=0.7\textwidth]{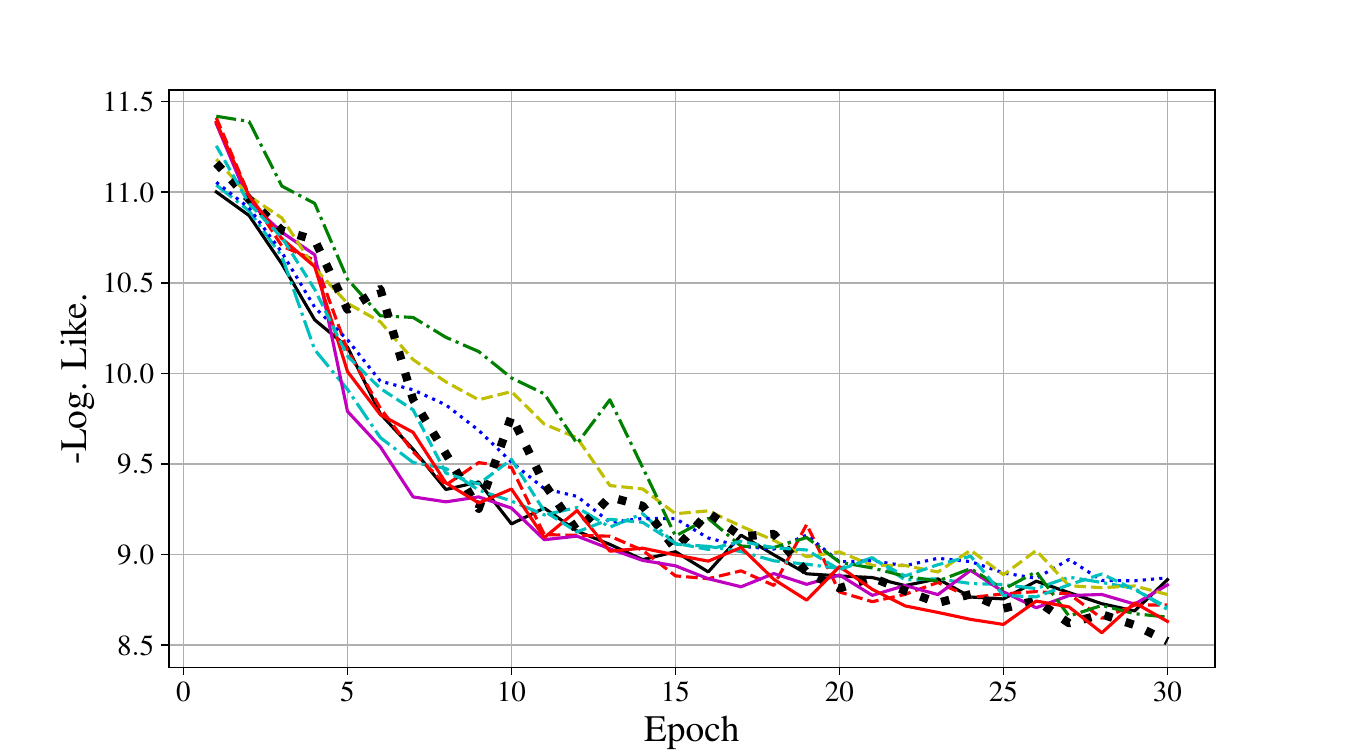}
    \includegraphics[width=0.28\textwidth]{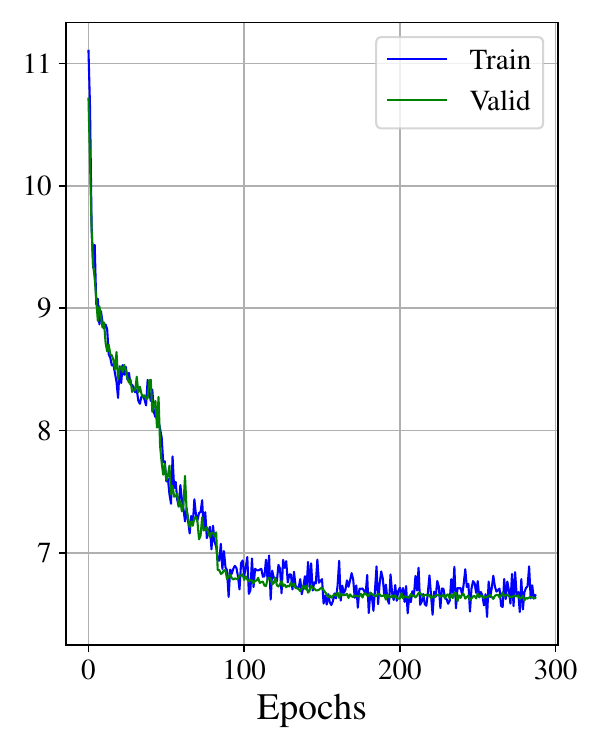}
    \caption{\textbf{Left}: Validation loss from the top 10 HPO runs in Table~\ref{tab:flow-hpo-table}.
    \textbf{Right}: Training/Validation loss for best model from the left panel until early-stopping.}
    \label{fig:flow-hpo-top-runs}
\end{figure}
Hyper-parameter optimization is done for $\sim 100$ configurations involving the
number of transforms, the configuration of each transform, learning rate, batch size, optimizer
weight decay and momentum parameters shown in Table~\ref{tab:flow-hpo-table} using asynchronous
hyper-banding with early stopping~\cite{li2018massively} using the {\raytune} library~\cite{liaw2018tune}.
This stops under-performing runs in favor of allowing better performing runs to continue. We carried
out the HPO runs over 4 A40 GPUs which took $\sim 15$ hours. The average validation
loss over validation epoch for the top 5 runs are shown in the left panel of Figure~\ref{fig:flow-hpo-top-runs}.
The right panel of the figure shows the training/validation loss of the best model configuration
trained until early-stopping condition is met.
While all the runs were allowed to run for 30 epochs, most of them are stopped early. The validation
loss for the top ten performing runs are shown in the figure.
We use the topmost configuration in Table~\ref{tab:flow-hpo-table} for the results of the paper. The training
and validation for this configuration, trained until early-stopping is shown on right panel of
Figure~\ref{fig:flow-hpo-top-runs}.

\section*{References}
\bibliographystyle{unsrt}
\bibliography{references}
\end{document}